\documentclass[amsmath,amssym,ams, preprint, superscriptaddress,nofootinbib]{article} 
\usepackage[a4paper,textwidth=17.9cm,textheight=26cm]{geometry}
\usepackage[english]{babel}
\usepackage[explicit]{titlesec}
\renewcommand{\baselinestretch}{1.5} 
\usepackage[T1]{fontenc}
\usepackage{lmodern}
\usepackage{helvet}
\usepackage{lineno}
\usepackage{graphicx}
\usepackage{xr}
\usepackage{authblk}
\usepackage[normalem]{ulem}
\usepackage{siunitx}
\usepackage{upgreek}
\usepackage{amsmath}

\newcommand{\nocontentsline}[3]{}
\newcommand{\tocless}[2]{\bgroup\let\addcontentsline=\nocontentsline#1{#2}\egroup}

\makeatletter
\newcommand*{\addFileDependency}[1]{
  \typeout{(#1)}
  \@addtofilelist{#1}
  \IfFileExists{#1}{}{\typeout{No file #1.}}
}
\makeatother

\addto\captionsenglish{}
\addto\captionsenglish{}

\begin{document}

\title{Coherent transduction and dynamical back-action between light and millimeter waves}
\author[1,2]{André Garcia Primo}
\author[1,2]{Jiawen Liu}
\author[1,2]{Amit Vankayala}
\author[1,2]{Gabriel Jülg}
\author[1,2]{Aleksei Gaier}
\author[1,2]{Yazan Lampert}
\author[1,2]{Ileana-Cristina Benea-Chelmus}

\affil[1]{Hybrid Photonics Laboratory, École Polytechnique Fédérale de Lausanne (EPFL), CH-1015, Switzerland}
\affil[2]{Center for Quantum Science and Engineering (EPFL), CH-1015, Switzerland}

\date{\today}
\maketitle

\begin{abstract}
    Electro-optic transduction can bypass the loss and noise of electronic frequency conversion at millimeter-wave frequencies by parametrically converting signals onto low-loss, low-noise optical carriers, with coupling rates that grow with frequency. Exploiting this growth, however, requires shrinking both the millimeter-wave and optical resonators while preserving their quality factors and spatial overlap, which has not yet been achieved. Here, we demonstrate a triply resonant transducer whose optical and millimeter-wave resonators have comparable dimensions and minimal mode volumes. Coupled lithium niobate Fabry--Pérot cavities, with free spectral ranges above \SI{650}{\giga\hertz} and quality factors up to $10^6$, form a photonic molecule integrated with a quarter-wave gold resonator at \SI{263}{\giga\hertz}. The device reaches a vacuum electro-optic coupling rate $g_0/2\pi \approx \SI{60}{\kilo\hertz}$, six times larger than previous millimeter-wave transducers and sixty times larger than at microwave frequencies. The device optically detects room-temperature thermal millimeter-wave photons at effective occupancies below one, which also serve as a calibration tone for transduction efficiency. Coherent transduction reaches cooperativities approaching $0.3$, corresponding to internal efficiencies around $67\,\%$, accompanied by electro-optic dynamical back-action that broadens or narrows the optical resonances. These results demonstrate the advantage and feasibility of the frequency scaling of electro-optic coupling, establishing an architecture for weak-signal sensing and quantum and classical networking.
\end{abstract}

\maketitle

\tocless\section{Introduction}\label{sec_intro}
Coherently converting signals between widely different frequencies is essential to both classical and quantum technologies. Classical communications must deliver high data rates over long distances, which makes optical carriers the natural choice. In next-generation networks, these carriers are modulated with signals that are synthesized and received in the millimeter-wave band (mm-wave, 30--300~GHz)~\cite{Rappaport2019,dang2020should}, at frequencies where electronic sources and detectors show a drop in performance~\cite{Imran2017,Zhao2026,Samoska2011,Siegel2002}. The mm-wave band is also attracting growing interest in quantum hardware. Rydberg transitions in atomic systems fall within this range, making it a natural band for interfacing atoms and superconducting circuits in hybrid quantum technologies~\cite{Kumar2023}. Higher operating frequencies further ease the thermal requirements of cryogenic systems. For instance, at \SI{100}{\giga\hertz} a quantum of energy corresponds to $\approx\SI{5}{\kelvin}$ and modes would remain near their ground states at temperatures roughly two orders of magnitude above the tens of millikelvin required at microwave frequencies~\cite{Anferov2020,Kumar2023,multani2026}. 
    Consequently, qubits have recently been demonstrated in the mm-wave band at temperatures up to  1~K~\cite{Huang2024,anferov2024_72GHz}, while dispersive readout of microwave qubits using high-frequency resonators~\cite{mencia2025,dixit2026} and microwave-to-mm-wave converters~\cite{Pechal2017,Hauer2021} promise to deliver the advantages of mm-waves without raising the frequency of every component in a quantum processing chain. Realizing quantum computation at such frequencies requires generating and detecting quantum-level mm-wave signals directly, which runs into the same limitations of electronic sources and detectors as classical systems. This has driven the development of frequency converters bridging the optical and mm-wave bands~\cite{Nagatsuma2016,Zhu2025,Li2023,Koenig2013}, such that mm-wave signals can be generated and detected with photonic components.

    Integrated electro-optic (EO) transducers are based on the nonlinear response of materials to coherently convert signals between the electric and optical domains through three-wave mixing~\cite{Han2021,Hu2025}. The optical domain is particularly attractive for routing the input and output signals in cryogenic systems, as optical fibers carry negligible heat loads and optical modes sit in their ground state even at room temperature. Although most demonstrations currently operate at microwave frequencies~\cite{Lecocq2021,Arnold2025,warner2025, Sahu2023}, extension to the mm-wave range in established materials such as thin-film lithium niobate (TFLN) and lithium tantalate is natural, as the Pockels coefficients that mediate the EO interaction are broadband and remain approximately constant up to the low-THz range, allowing a single platform to address the full 30--300~GHz band~\cite{liu2026, Chelladurai2025,lampert2024}. 
    
    The physics of such transducers also favors higher frequencies, because the vacuum EO coupling rate grows with the mm-wave frequency and inversely with the mm-wave mode volume, $g_0\propto\sqrt{\omega_\mathrm{mmw}/V_\mathrm{mmw}}$, while the mode volume itself shrinks with wavelength. For a resonator whose length scales as $1/\omega_\mathrm{mmw}$ at fixed transverse dimensions, $V_\mathrm{mmw}\propto1/\omega_\mathrm{mmw}$ and $g_0$ scales linearly with frequency~\cite{multani2026,Jzhang2026}, provided that the spatial overlap between the optical and mm-wave fields is preserved as the structure is scaled down. Moreover, the transduction efficiency is maximized when the cooperativity $C = 4 g_0^2 n_\mathrm{p}/(\kappa\gamma_\mathrm{mmw})$ reaches unity, where $n_\mathrm{p}$ is the intracavity pump photon number and $\kappa$ and $\gamma_\mathrm{mmw}$ are the optical and mm-wave loss rates, respectively. This frequency scaling therefore makes the mm-wave band attractive wherever the cooperativity is the constraint, as in the transduction of quantum states between superconducting or atomic processors, sensing, and optical networks. Large cooperativities further enable dynamical back-action, which modifies the damping rates of the interacting modes and allows control over their spectrum and phase. This gives access to tools known from Brillouin lasers, from pump-tunable filtering and amplification to self-oscillation with strong linewidth narrowing above $C=1$~\cite{Li2012,Aspelmeyer2014,Massel2011}.

    In practice, large cooperativities and dynamical back-action have so far remained out of reach in the mm-wave band. The main reason is that current systems do not meet the overlap condition on which the frequency scaling relies: the reduction in mm-wave mode volume with increasing frequency has not been accompanied by a similar reduction in the optical mode volume, leaving the two components of the transducer with disparate sizes and hence reduced spatial overlap. Indeed, while mm-wave resonators have reached wavelength scale, optical geometries remain constrained to racetrack resonators several millimeters long, with free spectral ranges in the 30--100~GHz range. Shrinking the optical dimensions is challenging~\cite{Gao2023}, as the stronger lateral confinement and tighter bending radii typical of these resonators increase radiative and sidewall-scattering losses~\cite{Khalatpour2026}. Furthermore, EO coupling rates stem mostly from resonator sections where the mm-wave and optical field polarizations couple to the large $r_{33}$ Pockels coefficient of TFLN~\cite{Hu2025,mckenna2020}, so structures containing bends have an intrinsically reduced useful interaction length. Consequently, gains from scaling the devices down are paid for with degraded optical quality factors and reduced overlap, with a net reduction in cooperativity. A direct comparison between the best reported microwave and mm-wave devices shows that an eighty-fold increase in transduction frequency (from $\approx4$ to $\approx310$~GHz) has yielded only an approximately tenfold increase in coupling rate~\cite{gaier2025,warner2025} (from $g_0/(2\pi)\approx 1$ to $\approx10$~kHz), an eightfold shortfall relative to the linear scaling expected above.

    Here, we address this problem by demonstrating a mm-wave-to-optical transducer based on a photonic molecule formed by lithium niobate Fabry--Pérot cavities, in which the optical and mm-wave resonators have comparable dimensions, preserving their spatial overlap at small mode volumes. The optical cavities combine free spectral ranges (FSRs) above \SI{650}{\giga\hertz} with quality factors up to $10^6$. Their hybridization generates a doublet of resonances split by $\approx\SI{263}{\giga\hertz}$, which allows them to be resonantly coupled to a quarter-wave gold resonator at the same frequency. The resulting vacuum coupling rate of $g_0/(2\pi) \approx \SI{60}{\kilo\hertz}$ is six times larger than in previous mm-wave transducers~\cite{gaier2025} and sixty times the rates reported at microwave frequencies~\cite{warner2025}, hence achieving the predicted linear scaling for $g_0$. This enables the optical readout of thermal mm-wave photons with an effective occupation around 0.2 photons, showing that mm-wave fields with occupations below one quantum can be resolved optically without amplification. We also reach cooperativities $C\approx 0.3$, which allow us to demonstrate coherent transduction with internal efficiencies of $\approx 67\,\%$, a range previously achieved only by hybrid photonic-atomic-superconducting devices~\cite{Kumar2023}, but now in an on-chip device, and observe electro-optic dynamical back-action in the mm-wave band for the first time, allowing dynamic control over optical and mm-wave lineshapes. These results are obtained in a regime where the mm-wave mode is almost two orders of magnitude more dissipative than the optical ones, in contrast to electro-optic and optomechanical systems at microwave frequencies~\cite{Mayor2025,Holzgrafe2020,Nunnenkamp2014}.

\tocless\section{Results}\label{sec_results}
\subsection*{Device design, working principles, and fabrication}

        \begin{figure*}[p!]
            \centering
            \includegraphics[width=17.0cm]{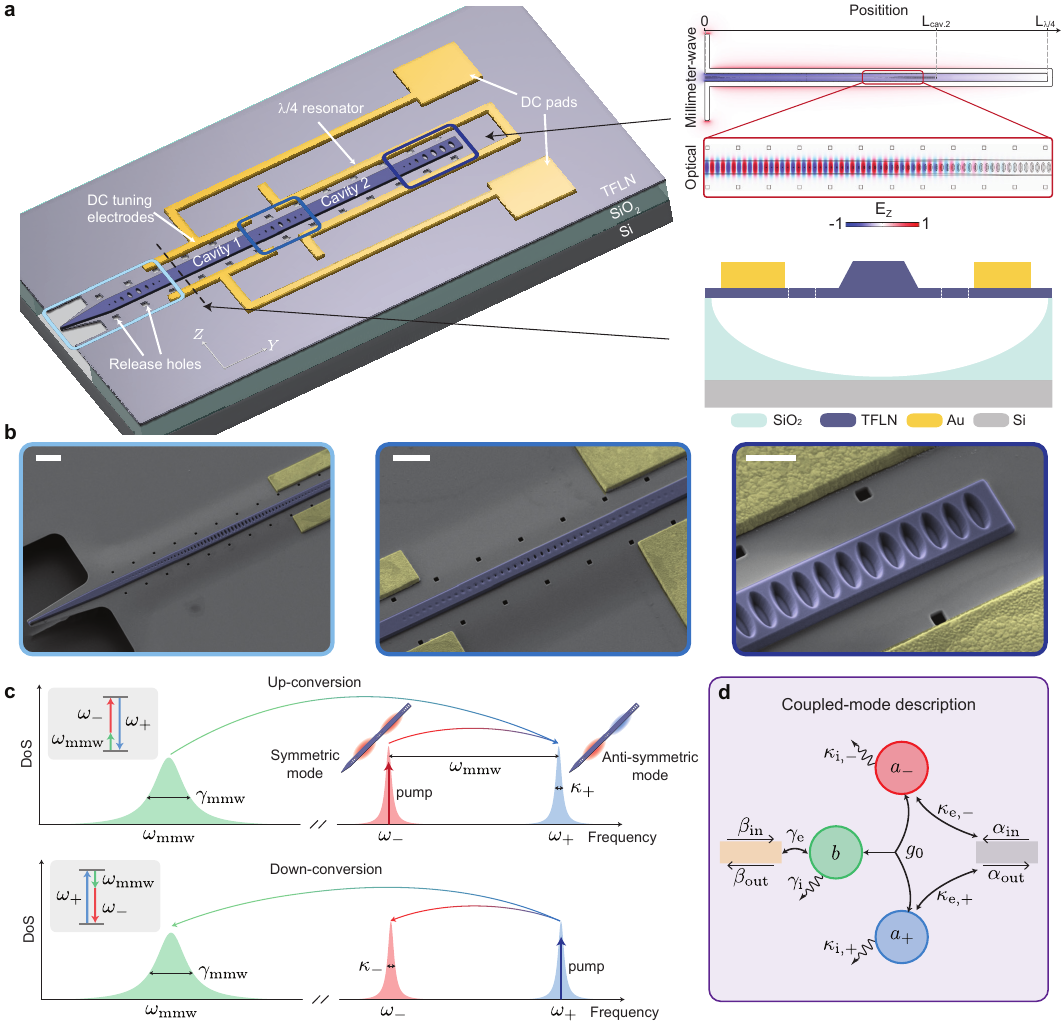}
            \renewcommand{\baselinestretch}{1} 
            \caption{\textbf{Triply resonant transducer based on lithium niobate Fabry--Pérot cavities.} 
            \textbf{a}, Device design. Fabry--Pérot cavities are evanescently coupled through a shared, partially transmissive photonic crystal mirror (blue box). Cavity 1 is addressed through an input mirror connecting it to a bus waveguide (light blue box), while cavity 2 is terminated by a highly reflective mirror (dark blue box). The mirrors are designed for reflectivities above 0.999 over a \SI{10}{\tera\hertz} band around the telecom C-band (see Supplementary Note 1). A gold quarter-wavelength ($\lambda/4$) millimeter-wave resonator is patterned around cavity 2 to provide electro-optic coupling, and DC tuning electrodes around cavity 1 to tune its resonance frequency. Release holes are patterned in the slab to enable oxide removal, with their distance from the optical structures chosen to not degrade optical quality factors. The device is designed such that the optical and mm-wave mode polarizations align to the crystal Z axis, maximizing electro-optic coupling. Top right: finite-element method simulations of the Z-polarized electric fields of the mm-wave and optical modes showing their spatial overlap. Cavity 2 is shorter than the quarter wavelength ($L_\text{cav,2}<L_{\lambda/4}$). Bottom right: cross-section of the device. The thin-film lithium niobate (TFLN) layer is suspended by removing the underlying $\mathrm{SiO_2}$ through the release holes. The millimeter-wave resonator and the tuning electrodes are both Au.
            \textbf{b}, False-colored scanning electron micrographs showing, from left to right, the input, shared and highly reflective mirrors with border colors matching the boxes in \textbf{a}. Gold is shown in yellow and the lithium niobate cavities in purple. Scale bars are \SI{3}{\micro\meter}, \SI{2}{\micro\meter}, and \SI{1}{\micro\meter}, respectively.
            \textbf{c}, Density of states (DoS) representation of the triply resonant transducer, illustrating the reversed dissipation regime $\gamma_\text{mmw} \gg \kappa_\pm$ and up- and down-conversion processes. Hybridization of cavities 1 and 2 yields the symmetric, $a_-$, and anti-symmetric, $a_+$, optical modes. Pumping $a_-$ at $\omega_-$ drives up-conversion whereas pumping $a_+$ at $\omega_+$ drives down-conversion. The optical frequency splitting is resonant with the millimeter-wave mode at $\omega_\text{mmw}$ to ensure triply resonant operation, $\omega_+ - \omega_- \approx \omega_\text{mmw}$.
            \textbf{d}, Coupled-mode description of the transducer. The optical modes $a_\pm$ have internal and external loss rates $\kappa_{\text{i},\pm}$ and $\kappa_{\text{e},\pm}$, and are coupled to the same waveguide with input and output amplitudes $\alpha_\text{in}$ and $\alpha_\text{out}$. The millimeter-wave mode $b$ is driven and read out with amplitudes $\beta_\text{in}$ and $\beta_\text{out}$ through an external channel at rate $\gamma_\text{e}$, and dissipates internally at rate $\gamma_\text{i}$. The optical modes are electro-optically coupled to $b$ with a rate $g_0$, which allows coherent conversion between them.
            }
            \label{fig:1}
        \end{figure*}

        The device concept is shown in Fig.~\ref{fig:1}\textbf{a}. Periodically patterned sections in a TFLN waveguide form photonic crystal mirrors that bound two coupled Fabry--Pérot cavities. Cavity 1 is excited through an input waveguide via a partially transmissive mirror. It has a total length of $\SI{110}{\micro\metre}$ and an FSR $\approx \SI{705}{\GHz}$, while cavity 2 is longer by $\SI{9}{\micro\meter}$ and has an FSR of
        $\approx\SI{655}{\giga\hertz}$. The two cavities are mutually coupled via a shared mirror, such that doublets of hybridized modes appear in the optical spectrum. The difference in FSRs ensures that the detuning between the modes of cavity 1 and 2 varies slowly across the spectrum and at least one mode pair lies close to degeneracy where hybridization is optimal. This gives rise to symmetric, $a_- \propto a_1+a_2$, and anti-symmetric, $a_+ \propto a_1-a_2$, hybrid modes, where $a_1$ and $a_2$ are the bare modes of cavities 1 and 2, respectively. The mm-wave component is a quarter-wavelength gold resonator with length $\approx\SI{165}{\micro\meter}$, designed to resonate at the splitting frequency of the maximally hybridized doublet, and patterned around cavity 2. A small antenna section (total length $\approx \SI{36}{\micro\metre}$) is incorporated into the resonator to allow its coupling to free-space mm-wave signals~\cite{Anferov2020}. DC tuning electrodes around cavity 1 tune its resonance electro-optically, compensating for residual detunings between the optical cavities and allowing the optical splitting to be matched to the mm-wave resonance frequency. The device is suspended to improve optical confinement and reduce mm-wave loss~\cite{Jiang2020}. Scanning electron micrographs of the fabricated devices (see Methods for fabrication details) are shown in Fig.~\ref{fig:1}\textbf{b}. Further details of the geometry are shown in Supplementary Note 1.  

         The full EO interplay is depicted in Fig.~\ref{fig:1}\textbf{d}. If an optical doublet is well isolated from other modes, the system can be described by a three-mode interaction Hamiltonian as $H = \hbar g_0(a_+ a_-^\dagger b^\dagger  + a_+^\dagger a_- b )$, where $a_+$, $a_-$, and $b$ are annihilation operators for the modes with frequencies $\omega_+$, $\omega_-$, and $\omega_\text{mmw}$, respectively. Such an interaction is triply resonant when $\omega_\text{mmw} = \omega_+ - \omega_-$. Placing an optical pump at mode $a_-$ leads to frequency up-conversion described by a beam-splitter interaction $H_\text{BS} = \hbar g_\text{EO}(a_+  b^\dagger  + a_+^\dagger b )$, promoting the coherent state transfer between $a_+$ and $b$. Pumping $a_+$ results in down-conversion and a two-mode squeezing Hamiltonian $H_\text{TMS} = \hbar g_\text{EO}(a_-^\dagger b^\dagger  + a_- b )$, allowing parametric interactions and the generation of entangled optical-mm-wave photon pairs, as shown in Fig.~\ref{fig:1}\textbf{c}. In both cases, the effective EO coupling is enhanced by the number of pump photons $n_{\text{p},\pm}$, as $g_\text{EO} = g_0 \sqrt{n_{\text{p},\pm}}$ (see Supplementary Note 2). In our system, the optical modes $a_\pm$ are coupled at rates $\kappa_{\mathrm{e},\pm}$ to a common waveguide, with input and output amplitudes $\alpha_\mathrm{in}$ and $\alpha_\mathrm{out}$, and have intrinsic loss rates $\kappa_{\mathrm{i},\pm}$. The mm-wave mode is coupled at a rate $\gamma_\mathrm{e}$ to its own channel, with input and output amplitudes
        $\beta_\mathrm{in}$ and $\beta_\mathrm{out}$, and has an intrinsic loss rate
        $\gamma_\mathrm{i}$. Importantly, in our system, the total optical losses $\kappa_\pm = \kappa_{\text{i,}\pm} + \kappa_{\text{e,}\pm}$ are much smaller than the total mm-wave losses $\gamma_\text{mmw} = \gamma_\text{i}+\gamma_\text{e}$, so we operate in the so-called reversed dissipation regime, $\kappa_\pm \ll\gamma_\text{mmw}$.
    
    \subsection*{Optical and millimeter-wave characterization}
    
        \begin{figure*}[t!]
        \centering
        \includegraphics[width=18cm]{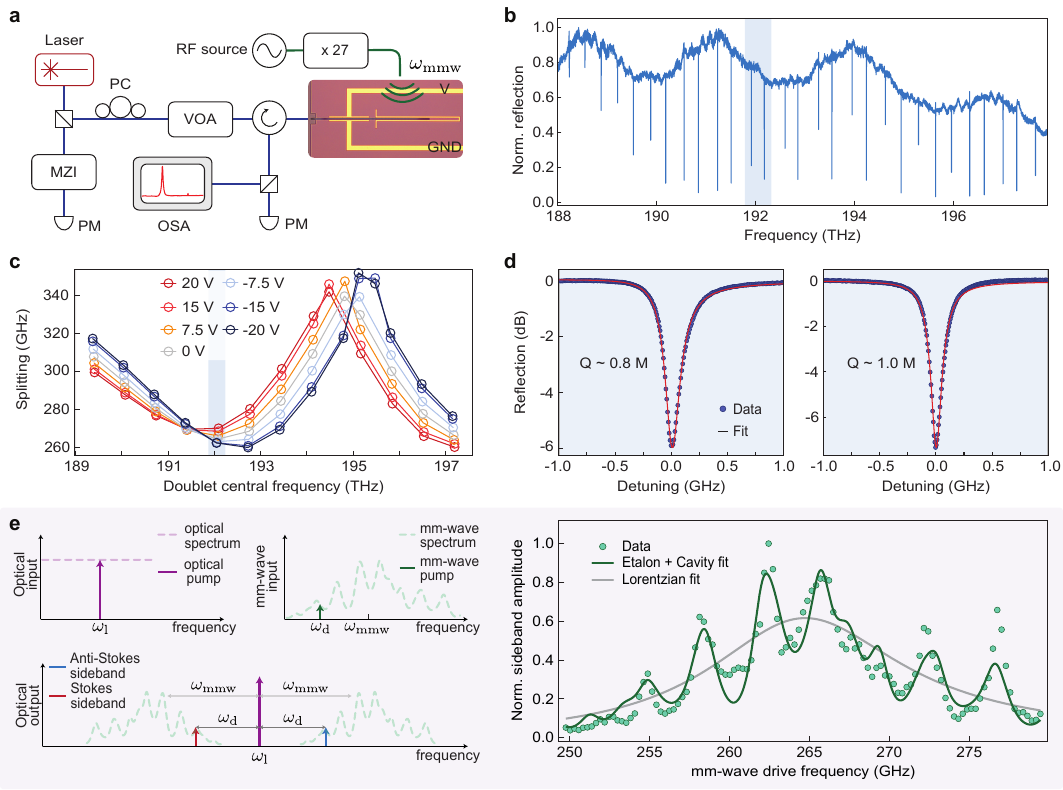}
        \renewcommand{\baselinestretch}{1} 
        \caption{\textbf{Optical and millimeter-wave characterization of the transducer.} 
        \textbf{a}, Experimental setup. A tunable laser is edge-coupled to the transducer through a lensed fiber, with a polarization controller (PC) matching the input polarization to that of the device and a variable optical attenuator (VOA) setting the input power. An optical circulator collects the reflected signal, which is split between a power meter (PM) and an optical spectrum analyzer (OSA). A Mach--Zehnder interferometer (MZI) of known free spectral range provides a relative frequency reference. Millimeter-wave signals are generated by a frequency multiplication chain ($\times 27$) that converts microwave tones near \SI{10}{\giga\hertz} to approximately \SI{270}{\giga\hertz}, and are radiated onto the chip from a waveguide taper.
        \textbf{b}, Normalized optical reflection spectrum, showing doublets of hybridized modes across the telecom band. The doublet used throughout is highlighted in blue.
        \textbf{c}, Electro-optic tuning of the doublet splitting for bias voltages between $-20$ and \SI{20}{\volt}. The tuning is linear with voltage such that the detuning $\delta$ between the modes of cavities 1 and 2 is proportional to the voltage $V$. When $\delta \ll 2J$, the doublet splitting stops changing appreciably. The observed splittings span \SIrange{260}{350}{\giga\hertz} and vary across the spectrum because the two cavities differ slightly in length, therefore changing $\delta$ across different mode pairs.
        \textbf{d}, Normalized reflection of the two modes of the doublet used in the experiments, as a function of detuning from resonance, with Lorentzian fits (red). The low- and high-frequency modes have quality factors of $\approx\num{0.8e6}$ and $\approx\num{1e6}$.
        \textbf{e}, Millimeter-wave spectrum characterization. The coupled cavities are replaced by a waveguide terminated in a photonic crystal mirror, so that the spectrum of a quarter-wave resonator with design identical to the transducer's can be sampled continuously in frequency space (see Methods). Left: The millimeter-wave and optical input tones interact electro-optically and generate sidebands. The amplitudes of the generated sidebands in the output reflected signal carry the information of the mm-wave spectrum. Right: Normalized sideband amplitude versus millimeter-wave drive frequency. Oscillations appear on top of a Lorentzian envelope (gray), arising from the interplay between the quarter-wave resonator and etalon modes of the silicon handle. A combined etalon and cavity fit (green solid curve) yields $\omega_\text{mmw}/2\pi = \SI{264\pm2}{\giga\hertz}$ and $\gamma_\text{mmw}/2\pi = \SI{15\pm4}{\giga\hertz}$.}
        \label{fig:2}
        \end{figure*}
    
         We characterize the optical and mm-wave responses of our transducer using the experimental setup in Fig.~\ref{fig:2}\textbf{a}, as described in Methods. The normalized optical reflection spectrum of our device is shown in Fig.~\ref{fig:2}\textbf{b}, where a sequence of doublets appears. For a given hybrid mode pair, the observed splittings $\omega_+ - \omega_-$ are set by the fabrication-defined coupling $J$ between the two bare modes of cavities 1 and 2 and their detuning $\delta$, i.e.~$\omega_+ - \omega_- = \sqrt{4 J^2 + \delta^2}$. Hence, by EO tuning one of the bare modes into (out of) resonance with its pair, one can effectively decrease (increase) the splitting, with a minimum of $2J$ where hybridization is complete~\cite{zhang2026}.
         The EO tuning in our device is shown in Fig.~\ref{fig:2}\textbf{c} for applied voltages between $-20$ and $\SI{20}{\volt}$, over which the splittings across different doublets span 260 to 350 GHz.
         Optimal operation requires $\delta \approx 0$ and consequently $\omega_+ - \omega_-\approx \SI{260}{\GHz}$, to ensure strong EO coupling. We therefore pick the optical modes shown in Fig.~\ref{fig:2}\textbf{d} for our experiments, which approach a splitting of $\SI{260}{\GHz}$ (close to the designed mm-wave frequency) for voltages around $\SI{-15}{\volt}$, as highlighted in blue in Figs.~\ref{fig:2}\textbf{b} and \textbf{c}. The chosen mode pair operates around 192 THz (1561 nm), well within the telecom C-band, displays quality factors $Q_- = 8.3\times 10^5$ ($\kappa_-/(2\pi)\approx \SI{230}{\MHz}$) and $Q_+=10^6$ ($\kappa_+/(2\pi)\approx \SI{186}{\MHz}$), which yield finesses between $2.9\times 10^3$ and $3.7\times 10^3$, substantially above the state-of-the-art in similar structures~\cite{Seo2026,Yang2025}. 
    
        We characterize the mm-wave cavity resonance frequency and quality factor optically using the technique in Ref.~\cite{gaier2025}, depicted in the left panel of Fig.~\ref{fig:2}\textbf{e}. The measurement is done on a device nominally identical to the transducer, but with the coupled cavities replaced by a simple waveguide with a continuous optical spectrum, as described in Methods. The resulting data is shown in the right panel of Fig.~\ref{fig:2}\textbf{e}. We observe oscillations on top of a Lorentzian response, in agreement with observations in similar systems~\cite{multani2026}. Their existence is attributed to substrate modes that dress the quarter-wavelength response, and hence, in our fitting, we lump such substrate modes into different etalon families that allow us to reproduce the observed behavior and estimate the mm-wave frequency and linewidth as $\omega_\text{mmw}/(2\pi) = \SI{264\pm2}{\GHz}$ and $\gamma_\text{mmw}/(2\pi) = \SI{15\pm 4}{\GHz}$, corresponding to a quality factor of $Q_\text{mmw} = \SI{17\pm4}{}$. In the following experiments we bias our optical cavities such that the doublet splitting closely matches the measured $\omega_\text{mmw}$. In the chosen mode pair, this requires only a residual detuning $\delta \ll 2J$, and therefore near-perfect hybridization is achieved. Further details on the mm-wave spectrum fitting process are described in Supplementary Note 3.

    \subsection*{Sensing of thermal photons}
    
        \begin{figure*}[t!]
        \centering
        \includegraphics[width=18cm]{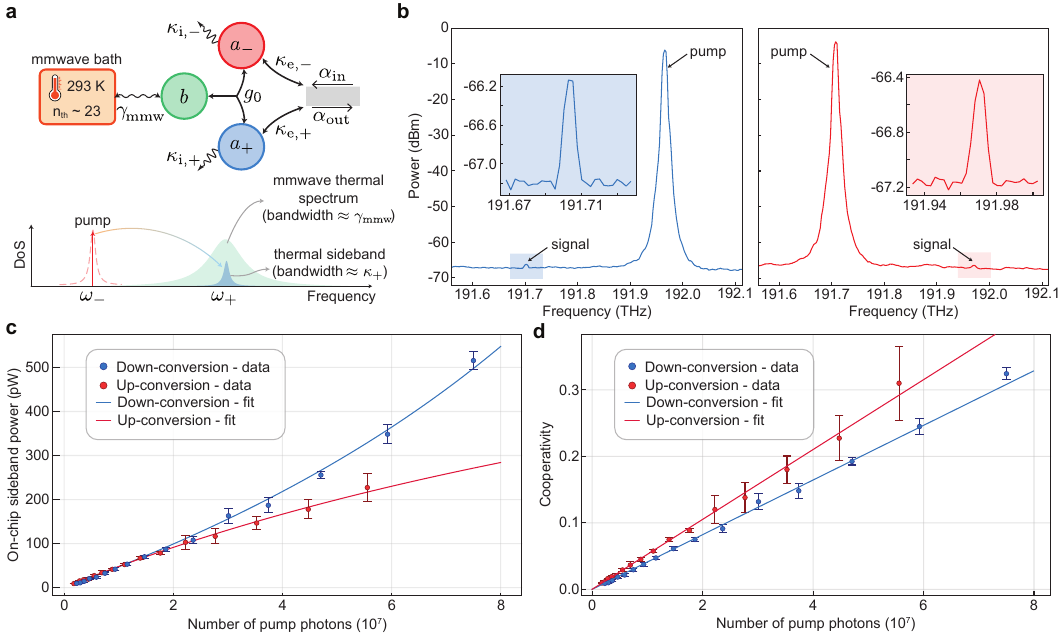}
        \renewcommand{\baselinestretch}{1} 
        \caption{\textbf{Optical measurement of millimeter-wave thermal photons.} 
        \textbf{a}, Working principle of the transducer when the millimeter-wave mode $b$ is driven by thermal fluctuations at room temperature. The thermal occupation of $b$ seeds sideband generation when either of the optical modes $a_\pm$ is pumped. Because the total occupation $n_\text{th} \approx 23$ is spread over the millimeter-wave bandwidth $\gamma_\text{mmw} \gg \kappa_\pm$, the transducer samples only the fraction falling within the linewidth of the optical mode receiving the converted photons, giving an effective transduced occupation of $2 n_\text{th}\kappa_\pm/(\pi\gamma_\text{mmw}) = \num{0.21\pm0.08}$, below the single-photon level. Bottom panel: density of states (DoS) illustrating the bandwidth mismatch. 
        \textbf{b}, Optical spectra for down-conversion (left) and up-conversion (right), with the thermally generated sidebands highlighted and magnified in the insets. Traces were recorded at intracavity pump photon numbers $n_{\text{p},-} = \num{2.8\pm0.1e7}$ and $n_{\text{p},+} = \num{3.7\pm0.1e7}$.
        \textbf{c}, On-chip sideband power versus intracavity pump photon number for both down-conversion and up-conversion processes. Both grow linearly at low power and under stronger pumping down-conversion becomes superlinear and up-conversion sublinear, reflecting the cooperativity dependences $C/(1-C)$ and $C/(1+C)$. Fits to the two datasets yield consistent values of $g_0/2\pi = \SI{60\pm8}{\kilo\hertz}$. 
        \textbf{d}, Cooperativity extracted point by point, independently of $g_0$ and of the detection calibration. Both branches scale linearly with pump photon number and reach $C \approx 0.3$ at the highest powers. Error bars in \textbf{c} and \textbf{d} represent one standard deviation across averaged optical spectra.
        }
        \label{fig:3}
        \end{figure*}

        Now we use our transducer to probe room-temperature mm-wave thermal signals and show that we can resolve them with sub-photon effective occupations. Operating our device at room temperature in the absence of any coherent drive yields a purely thermally populated mm-wave mode, with an occupation  $n_\text{th}\approx 23$ at a temperature of 293 K, given by the Bose--Einstein distribution. These photons are continuously distributed across the mm-wave lineshape, roughly spanning the mm-wave linewidth $\gamma_\text{mmw}$. Pumping one of the optical modes of our system therefore generates up- and down-converted optical sidebands driven by the thermal occupation. The bandwidth of the generated signal is limited by the optical linewidth, which is $\kappa_\pm/(2\pi)\approx \SI{200}{\MHz}$, roughly two orders of magnitude lower than $\gamma_\text{mmw}/(2\pi) = \SI{15\pm 4}{\GHz}$. Therefore our transducer can only sample the fraction of the mm-wave thermal photons that falls within the optical bandwidth, yielding an effective occupation of $\approx 2\, n_\text{th} \kappa_\pm/(\pi\gamma_\text{mmw}) = \SI{0.21\pm 0.08}{}$, below the single-photon level (see Supplementary Note 2). This mechanism is illustrated in Fig.~\ref{fig:3}\textbf{a}.
    
        In Fig.~\ref{fig:3}\textbf{b}, we show the reflected spectra of our transducer when down-conversion (left) and up-conversion (right) are seeded by the thermal occupation of the mm-wave cavity. The spectra are recorded directly with an optical spectrum analyzer, rather than through the electronic heterodyne detection used at lower frequencies~\cite{Jzhang2026}, which becomes costly and noisy above \SI{100}{\giga\hertz}. Despite the low effective thermal occupations probed by the transducer, we observe sidebands occurring at approximately $ \SI{263}{\GHz}$ frequency separation from the pump, matching the optical splitting. We record the on-chip sideband power as a function of intracavity pump photon number, as shown in Fig.~\ref{fig:3}\textbf{c}. We observe a linear trend for both up- and down-conversion at low input powers, and superlinear (sublinear) behavior for down-conversion (up-conversion) under strong pumping because of dynamical back-action, described later in the text. Such behavior is consistent with theoretical predictions for a sideband arising from thermal mm-wave photons in the reversed dissipation regime with sufficiently large cooperativities, as described in detail in Supplementary Note 2.

        The direct electro-optical detection of thermal photons above 100 GHz opens the possibility of using its signal as an absolute calibration reference for characterizing the transducer performance. Using the room temperature thermal occupation, fits to both up- and down-conversion datasets yield consistent values $g_0/(2\pi) = \SI{60\pm8}{\kHz}$. The cooperativity can also be obtained, without reference to $g_0$, by solving the sideband amplitude model for $C$ point by point (see Supplementary Note 2), as shown in Fig.~\ref{fig:3}\textbf{d}. We obtain $C\approx0.3$ for the highest pump powers in our experiments, and observe a linear increase with the number of photons, as expected. The resulting curves display distinct slopes because $C\propto1/\kappa_\pm$ and the two optical modes of our doublet display different loss rates. Operating the devices under larger input powers than those reported here led to a sudden drop in optical quality factor, which will be the subject of future investigation.
    
        \subsection*{Coherent transduction and back-action}
    
        \begin{figure*}[t!]
        \centering
        \includegraphics[width=18cm]{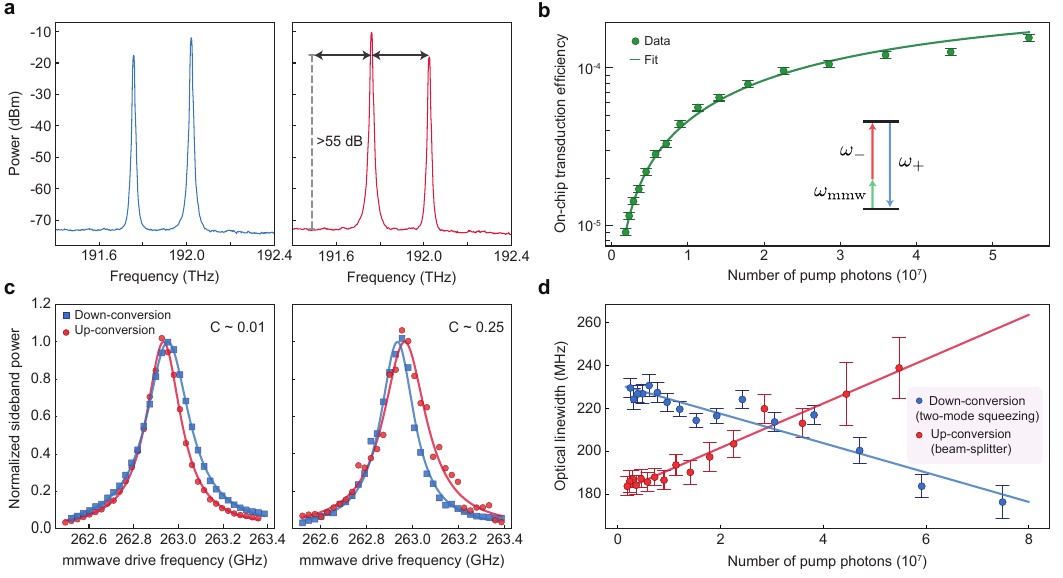}
        \renewcommand{\baselinestretch}{1} 
        \caption{\textbf{Coherent transduction and back-action measurements.} 
        \textbf{a}, Reflected optical spectra for down-conversion (left) and up-conversion (right), with the pump appearing as the high-power peak. The millimeter-wave source power was set to $P^\text{in}_\text{mmw} = \SI{2.2\pm0.1}{\milli\watt}$, measured with a calorimetric power meter, at intracavity pump photon numbers $n_{\text{p},+} = \num{7.7\pm0.2e6}$ and $n_{\text{p},-} = \num{5.7\pm0.2e6}$. The non-resonant sideband is suppressed by more than \SI{55}{\decibel} in both cases, limited by the noise floor of the detection. 
        \textbf{b}, On-chip millimeter-wave-to-optical transduction efficiency versus intracavity pump photon number. The source power is attenuated to $P^\text{in}_\text{mmw} = \SI{0.22\pm0.01}{\milli\watt}$ to avoid nonlinear frequency shifts of the sideband mode as the millimeter-wave frequency is swept. Efficiencies reach $\num{1.6\pm0.1e-4}$ for $n_{\text{p},-} > \num{5e7}$, corresponding to $C = \num{0.27\pm0.03}$. A fit to the data yields $g_0/2\pi = \SI{60\pm8}{\kilo\hertz}$, consistent with the thermal-sideband analysis. Inset: level diagram of the up-conversion process.
        \textbf{c}, Normalized sideband power versus millimeter-wave drive frequency for both processes, at low ($C \approx 0.01$, $n_{\text{p},+} = \num{6.1\pm0.2e6}$, $n_{\text{p},-} = \num{4.5\pm0.2e6}$) and high cooperativity ($C \approx 0.25$, $n_{\text{p},+} = \num{7.5\pm0.2e7}$, $n_{\text{p},-} = \num{5.5\pm0.2e7}$), measured through the optical linewidth variation. These traces probe the transduction bandwidth, set by the optical linewidth of the sideband mode, which broadens for up-conversion and narrows for down-conversion as the cooperativity increases.
        \textbf{d}, Optical linewidth of the sideband modes versus intracavity pump photon number, showing the linear broadening and narrowing expected from dynamical back-action. Linear fits yield $g_0/2\pi = \SI{50\pm7}{\kilo\hertz}$ for down-conversion and $\SI{63\pm9}{\kilo\hertz}$ for up-conversion. Error bars in \textbf{b} and \textbf{d} represent one standard deviation given by the fit uncertainty.
        }
        \label{fig:4}
        \end{figure*}
    
        We next characterize the transduction of coherent mm-wave signals, measuring its on-chip and internal efficiencies, its bandwidth, and the associated dynamical back-action. Because the OSA resolution prevented us from extracting the bandwidth in the previous measurements, we now resonantly pump one optical mode while sweeping the frequency of a Hz-linewidth mm-wave drive~\cite{gaier2026} and record the reflected optical spectrum, so that the sideband amplitude versus drive frequency traces out the transduction bandwidth. Transduction efficiency peaks when the drive frequency matches the optical splitting, i.e.~the triply resonant condition (see Supplementary Note 2). Examples for down-conversion (left) and up-conversion (right) are shown in Fig.~\ref{fig:4}\textbf{a}. Because the doublets are well isolated, the off-resonant sideband is suppressed by more than 55~dB relative to the resonantly enhanced one (limited by the noise floor of our detection chain), so the device also operates as a single-sideband modulator.

        We define the up-conversion efficiency as the ratio of the out-coupled sideband photon flux to the input mm-wave photon flux, $\eta = \dot{N}^\text{out}_\text{opt}/\dot{N}^\text{in}_\text{mmw}$, and observe values up to $\eta = \SI{1.6\pm0.1}{\times 10^{-4}}$ (Fig.~\ref{fig:4}\textbf{b}). The optical flux is referenced to the chip output, with fiber-coupling and propagation losses to the OSA calibrated out, whereas the mm-wave flux is referenced to the source output. The efficiency $\eta$ therefore also includes the coupling of the free-space mm-wave tone into the antenna and onto the quarter-wavelength resonator, which we model together as a single lumped resonator. Under triply resonant conditions, $\eta = (\gamma_\text{e}/\gamma_\text{mmw})(\kappa_{\text{e},+}/\kappa_+)\,4C/(1+C)^2$~\cite{Han2021}. We measure the optical extraction ratio $\kappa_{\text{e},+}/\kappa_+ = \SI{0.28\pm0.02}{}$ independently from the optical spectrum, and obtain $\gamma_\text{e}/\gamma_\text{mmw} = \SI{8.5\pm0.6}{\times10^{-4}}$ by using the thermal sidebands as a calibration tone of known occupation, which yields the intracavity mm-wave photon number under coherent drive without requiring knowledge of $g_0$ (see Supplementary Note 2). Together, these ratios set the maximum attainable efficiency, $\eta_\text{max} = \SI{2.4\pm0.2}{\times10^{-4}}$ at $C=1$, limited mainly by the source-to-chip mm-wave coupling. The measured efficiency thus corresponds to a maximum internal efficiency $4C/(1+C)^2 \approx 0.67$, a cooperativity $C = \SI{0.27\pm0.03}{}$, and a fitted $g_0/(2\pi) = \SI{60\pm8}{\kHz}$, in agreement with the thermal sideband analysis.

        Figure~\ref{fig:4}\textbf{c} shows the sideband amplitude versus mm-wave frequency for down- and up-conversion at low ($C\approx 0.01$) and high ($C\approx 0.25$) cooperativity. At low $C$, Lorentzian fits yield up- and down-conversion bandwidths that match the optical linewidths $\kappa_+$ and $\kappa_-$, respectively, obtained from our optical characterization. At high $C$, dynamical back-action dresses these linewidths through the interaction with the more dissipative mm-wave mode. In up-conversion, the beam-splitter interaction opens an additional decay channel for $a_+$ through the lossy mm-wave mode, broadening its optical linewidth to $\kappa_{\text{eff},+} = \kappa_+(1 + C)$. Conversely, in down-conversion, the two-mode squeezing interaction anti-damps $a_-$, narrowing its linewidth to $\kappa_{\text{eff},-} = \kappa_-(1 - C)$ (see Supplementary Note 2 for the derivation). Comparing the data at high cooperativity for down- and up-conversion, we verify that indeed the linewidths of $a_+$ and $a_-$ are pushed in opposite directions, to the extent that $a_+$ becomes more lossy than $a_-$, i.e.~$\kappa_{\text{eff},+} > \kappa_{\text{eff},-}$, in contrast to the data at low $C$, where $\kappa_+<\kappa_-$. Power-resolved measurements are shown in Fig.~\ref{fig:4}\textbf{d}, displaying linear and opposite trends for up- and down-conversion. Fits to the data allow us to extract $g_0/(2\pi) = \SI{50\pm7}{\kHz}$ for down-conversion, and $g_0/(2\pi) = \SI{63\pm9}{\kHz}$ for up-conversion, consistent within uncertainty with the previously independently extracted values. Besides providing an independent calibration of the EO coupling rates and cooperativities, the observation of dynamical back-action shows that the EO interaction can be used to reshape the optical spectrum in situ, for instance to switch an optical resonance between under- and overcoupling.

\tocless\section{Discussion}

    By matching the dimensions of the optical and mm-wave resonators, we demonstrated that the linear frequency scaling of the electro-optic coupling, which had previously been elusive, is achievable. The transducer is able to read out its own thermal environment at the sub-photon level, establishing that the optical readout of quantum-level mm-wave cavity fields is possible. Additionally, simple changes to the length of the resonators adapt the device for operation at different frequencies, which together with the flat electro-optic response of lithium niobate, positions the architecture as a route to quantum transduction and high-efficiency wireless optical interfaces for the sub-terahertz bands targeted by 6G. 

    The on-chip wireless transduction efficiency of $1.6\times10^{-4}$ is limited by the coupling of the free-space beam into these particular antenna and quarter-wave resonator designs, $\gamma_\text{e}/\gamma_\text{mmw} = \num{8.5e-4}$, rather than by the internal efficiency of 67\%. This could be addressed by optimizing the device antenna design and using silicon and standard dielectric lenses for mm-waves to improve the coupling tenfold~\cite{gaier2025}. Such an approach would come at the expense of a $\approx1/\sqrt{2}$ reduction in $g_0$, and lead to a predicted efficiency of $\approx 10^{-3}$. Alternatively, mm-wave signal delivery could be done directly to the resonator by using RF probes, in which case critical coupling could be achieved~\cite{Jzhang2026}, bringing efficiencies close to 10\%. More substantially, our fabrication process is compatible with superconducting films such as NbTiN or NbN~\cite{Frasca2023}, whose superconducting gap lies well above \SI{263}{\giga\hertz}. Replacing the gold resonator with a superconductor would reduce $\gamma_\text{i}$ by more than an order of magnitude~\cite{multani2026}, raising $\gamma_\text{e}/\gamma_\text{mmw}$ and the cooperativity at fixed pump power in equal measure, translating directly into transduction efficiency. At such high frequencies, ground-state operation even at temperatures around 4 K can be envisioned. 

    Lastly, to our knowledge, dynamical back-action has not previously been observed in electro-optic transducers in the mm-wave band, where the reversed dissipation hierarchy demands coupling rates high enough to overcome the large mm-wave losses. At the cooperativities demonstrated here, back-action already provides a tuning knob for the optical damping rate, as observed in our experiment. This allows, for example, resonances to be moved in and out of critical coupling and hence our device acts as a reconfigurable optical filter and delay line. In the mm-wave domain, the same interaction produces a response analogous to electromagnetically induced transparency and absorption~\cite{Agarwal2010}, with direct applications to mm-wave signal processing~\cite{Fleischhauer2005,Massel2011}. Superconducting electrodes could raise the cooperativity above $C = 1$, where down-conversion crosses the parametric instability threshold and the system displays self-sustained oscillations. In the reversed dissipation regime it is the optical sideband that oscillates, spectrally narrowed relative to the pump, as in Brillouin lasers~\cite{Li2012}. Raising the mm-wave quality factor beyond $\approx 2\times10^3$ instead restores the normal dissipation regime, in which back-action acts on the mm-wave mode rather than on the optical one~\cite{Qiu2023}. Crossing the same threshold in this regime would make an optically pumped transducer emit coherent mm-wave radiation, offering a route to efficient chip-scale mm-wave sources driven entirely by light.

\tocless\section{Methods}
   \subsection*{Fabrication}\label{methods:fab}

        The devices were fabricated on X-cut TFLN (NanoLN), initially consisting of a \SI{600}{\nano\meter} TFLN layer on a \SI{4.7}{\micro\meter} buried oxide with a $\approx\SI{500}{\micro\meter}$ thick high-resistivity silicon handle. The TFLN layer was trimmed by ion-beam etching (IBE) to a thickness of \SI{400}{\nano\meter}. The photonic cavities were patterned using a \SI{22}{\percent} dilution HSQ hard mask (EM Resist Ltd.) and electron-beam lithography (EBL), followed by development in \SI{25}{\percent} TMAH. These patterns were transferred into the TFLN layer by IBE to a depth of \SI{300}{\nano\meter}, leaving a \SI{100}{\nano\meter} slab. A second EBL step using undiluted ZEP defined the release holes, which were developed in $n$-amyl acetate, rinsed in MIBK:IPA (90:10), and etched by IBE. Amorphous lithium niobate redeposition was subsequently removed with a combination of warmed \SI{40}{\percent} KOH and RCA-1 solution, and the devices were annealed at \SI{520}{\celsius} for \SI{2}{\hour} in an oxygen atmosphere.
        A third EBL exposure defined the gold structures using a PMMA/MMA resist stack for lift-off. A \SI{7}{\nano\meter} Cr adhesion layer followed by \SI{300}{\nano\meter} Au was evaporated, and lift-off was completed in acetone. The chip was then diced to provide optical access to the edge couplers. Finally, the devices were released in buffered oxide etch (7:1) until an undercut of $\approx\SI{4}{\micro\meter}$ was obtained.

    \subsection*{Experimental setup and measurements} \label{methods:setup}
        A tunable laser (Toptica CTL 1550) is coupled in and out of the chip using a lensed fiber (OZ Optics TSMJ-3A-1550-9/125-0.25-7-2.5-14-2-AR) with $\approx 40\%$ coupling efficiency. A variable optical attenuator (OZ Optics DA-100-3A-1300/1550-9/125-S-60) and an optical circulator (Thorlabs 6015-3-APC) are used to control input power levels and retrieve the signal reflected by the device, respectively. This signal is then split and sent to an optical spectrum analyzer (Anritsu MS9740B) and a power meter (Thorlabs S122C) for optical characterization and spectroscopy. The relative wavelength of the laser is monitored using a fiber Mach--Zehnder interferometer (built in-house using two Thorlabs TW1550R5A2 50:50 beam splitters) with an independently calibrated FSR of $\SI{278 \pm 5}{\MHz}$, allowing the characterization of optical quality factors with high resolution. Electro-optic tuning is done through a signal-ground RF probe (|Z| probe Z04-B-GS-500) contacting on-chip pads, driven by a voltage source (Keithley 2400 Series SourceMeter). Finally, the mm-wave excitation employs a frequency multiplier spanning 240--380 GHz (Virginia Diodes WR-9.0 SGX + WR-2.8 frequency tripler), which is driven by a microwave tone generated using a vector network analyzer (Anritsu ShockLine MS46522B). The signal is radiated into free-space via a WR 2.8 waveguide taper which drives the chip from its back side. The alignment to the quarter-wavelength resonators is done with feedback from transduction measurements described in the text. Finally, mm-wave power calibrations were performed using a calorimetric power meter (Virginia Diodes PM5B).

    \subsection*{Millimeter-wave spectrum characterization}
         The light from an excitation laser is phase modulated by a mm-wave tone with varying frequency, and the amplitude of the generated sidebands yields information about the mm-wave spectrum, which can be characterized using the same setup in Fig.~\ref{fig:2}\textbf{a}. In coupled cavities, this technique is limited by their discrete spectrum, which samples the mm-wave response only at isolated frequencies. Therefore, we pattern devices next to our transducers where the coupled-cavity system is substituted by a simple waveguide with a photonic crystal mirror at its end, leaving the quarter-wave resonator, electrode geometry and release holes nominally identical. The on-chip proximity ensures that the transducer and the waveguide see similar fabrication conditions, such that the mm-wave response remains reasonably similar between devices. We optically pump the waveguide around 1560 nm while a mm-wave tone drives the quarter-wavelength resonator directly from free-space from the back side of the chip. The amplitudes of the optical sidebands are measured as a function of the mm-wave frequency $\omega_\text{mmw}$ using the OSA.

\vspace{1cm}

\textbf{Data Availability}
    The measurement data generated in this study will be deposited in the Zenodo database under 10.5281/zenodo.23811476.

\medskip
\textbf{Code Availability}
    The codes used to plot the data within this paper will be deposited in the Zenodo database under 10.5281/zenodo.23811476.

\medskip
\textbf{Acknowledgments}
    The authors acknowledge Fatemeh Arefi, Laetitia Imstepf, Zdenek Benes, and Makhlad Chadid for useful discussions and assistance with fabrication.
    AGP acknowledges funding through the Excellence Postdoctoral Fellowship Programme from the Quantum Science Center at EPFL and the Swiss National Science Foundation grant number 515449 (SPARK). JL acknowledges funding through the Excellence Postdoctoral Fellowship Programme from the Quantum Science Center at EPFL and the Swiss National Science Foundation grant number 503051 (Swiss Quantum Call). GJ acknowledges support from the Swiss National Science Foundation through grant number 515075 (SNSF-NSF lead agency project). AG acknowledges funding from the European Union’s Horizon Europe research and innovation programme under project MIRAQLS with grant agreement No. 101070700 and funding from the Swiss National Science Foundation Grant No. 219406. YL acknowledges funding from the European Union’s Horizon Europe research and innovation program under project MIRAQLS with grant agreement No 101070700. The chips were fabricated at the Center of MicroNanoTechnology (CMi) at EPFL.

\medskip
\textbf{Author contributions}
    AGP and ICBC conceived the idea; AGP conceived, designed, and fabricated the devices, performed the experiments, and analyzed the data; JL, AV, and GJ assisted with fabrication; JL, AG, and YL assisted with measurements and data analysis; AGP wrote the manuscript with input from all authors; ICBC supervised the project.
    
\medskip
\textbf{Competing interests}
    AGP, AG and ICBC are inventors on European patent application EP26215332.3, filed by EPFL and currently pending, which covers the use of Fabry–Pérot cavities for millimeter-wave electro-optic coupling, as described in this work. The remaining authors declare no competing interests.

\medskip
\textbf{Corresponding authors}
    Correspondence to André Garcia Primo (andre.garciaprimo@epfl.ch) or Ileana-Cristina Benea-Chelmus (cristina.benea@epfl.ch).

\appendix
\setcounter{section}{0}
\renewcommand{\thesection}{\arabic{section}}
\titleformat{\section}{\normalfont\Large\bfseries}{Supplementary Note \thesection.}{0.5em}{#1}
\setcounter{table}{0}
\renewcommand{\thetable}{S\arabic{table}}
\setcounter{figure}{0}
\renewcommand{\thefigure}{S\arabic{figure}}
\setcounter{equation}{0}
\renewcommand{\theequation}{S\arabic{equation}}
\renewcommand{\thefigure}{S\arabic{figure}}
\clearpage 
\vspace*{1cm}

\begin{center}
    {\textbf{\Large Supplementary information for} \\ \Large{Coherent transduction and dynamical back-action between light and millimeter waves}}
\end{center}

\date{}
\author{
\parbox{\textwidth}{
\centering
André Garcia Primo, Jiawen Liu, Amit Vankayala, Gabriel Jülg,\\ Aleksei Gaier, Yazan Lampert, and Ileana-Cristina Benea-Chelmus.
}
}

\vspace{1cm}
\tableofcontents
\newpage
\section*{Summary of device parameters}
\begin{table}[h!]
\caption{\textbf{Device parameters.} The four determinations of the vacuum electro-optic coupling
rate $g_0$ are obtained from different measurements and agree within their
uncertainties. $\mu_\text{fiber-to-chip}$ refers to the fiber-to-chip coupling efficiency.}
\label{tab:params}
\centering
\begin{tabular}{ccc}
\hline
Parameter & Value & Method \\
\hline
$\kappa_{-}/2\pi$                       & \SI{230}{\mega\hertz}          & Reflection measurement \\
$\kappa_{\mathrm{e},-}/2\pi$            & \SI{57}{\mega\hertz}           & Reflection measurement \\
$\kappa_{+}/2\pi$                       & \SI{186}{\mega\hertz}          & Reflection measurement \\
$\kappa_{\mathrm{e},+}/2\pi$            & \SI{53}{\mega\hertz}           & Reflection measurement \\
$(\omega_+-\omega_-)/(2\pi)$                                   & \SI{262.93\pm0.06}{\giga\hertz} & Coherent transduction vs.\ mm-wave frequency \\
$\text{mean FSR}$                      & \SI{681\pm 5}{\giga\hertz}    & Broadband transmission scan \\
$\omega_\mathrm{mmw}/2\pi$              & \SI{264\pm2}{\giga\hertz}      & Sideband response vs.\ mm-wave frequency \\
$\gamma_\mathrm{mmw}/2\pi$              & \SI{15\pm4}{\giga\hertz}       & Sideband response vs.\ mm-wave frequency \\
$\gamma_\mathrm{e}/2\pi$                & \SI{13\pm4}{\mega\hertz}       & Thermal sideband as calibration tone \\
$g_0/2\pi$                              & \SI{60\pm8}{\kilo\hertz}       & Thermal sideband analysis \\
$g_0/2\pi$                              & \SI{60\pm8}{\kilo\hertz}       & Coherent transduction \\
$g_0/2\pi$                              & \SI{50\pm7}{\kilo\hertz}       & Dynamical back-action (down-conversion) \\
$g_0/2\pi$                              & \SI{63\pm9}{\kilo\hertz}       & Dynamical back-action (up-conversion) \\
$\mu_\text{fiber-to-chip}$                              & \SI{0.4\pm0.02}{}      &Reflection measurement \\
\hline
\end{tabular}
\end{table}

\section{Device design}

\subsection{Fabry--Pérot cavities geometry}

    Our Fabry--Pérot cavities are formed between two photonic-crystal mirrors. The mirror unit cell is defined by four parameters: the width of the patterned thin-film lithium niobate (TFLN) waveguide, measured at its top surface and thus excluding the angled sidewalls; the lattice constant, defined as the center-to-center distance between consecutive elliptical holes; and the two semi-axes of the etched elliptical holes. To mitigate scattering losses caused by abrupt changes in the effective index between adjacent cells, these parameters are varied adiabatically from the bare-waveguide geometry, which contains no holes, to the target unit-cell geometry. This transition is parametrized by~\cite{Mayor2025}
    \begin{equation}
    p_n = p_{\infty}-\left(p_{\infty}-p_0\right)2^{-\left(n/\delta\right)^{M}},
    \end{equation}
    where $p_n$ denotes any one of the four unit-cell parameters defined above, evaluated at the $n$-th unit cell of a mirror. The quantities $p_0$ and $p_\infty$ are the initial and asymptotic values of the parameter, corresponding to $n = 0$ and $n \to \infty$, respectively. The constants $M$ and $\delta$ control how smoothly $p_n$ varies with $n$ and determine the number of unit cells after which $p_n$ no longer changes appreciably. Specifically, $\delta$ is the cell index at which $p_n$ reaches the midpoint between $p_0$ and $p_\infty$, while $M$ sets the steepness of the transition.

    Both ends of the input and central mirrors are connected to bare waveguides. The adiabatic transition is therefore applied twice: the parameters are first tapered from the bare-waveguide geometry to the target unit-cell geometry and then tapered back to the bare-waveguide geometry, resulting in a partially reflecting mirror of finite length. The two transitions are symmetric about the center. In contrast, the end mirror is connected to a bare waveguide on one side only and therefore contains a single transition. The input mirror is engineered to yield a coupling quality factor of the same order as the internal quality factor of the cavities, such that the pumping of our resonances is efficient. The central mirror is engineered to allow hybridization of the optical modes of cavity 1 and 2 such that the resulting splitting matches the mm-wave resonator frequency around $\SI{260}{\GHz}$.

\begin{figure}[t!]
            \centering
            \includegraphics[width=18cm]{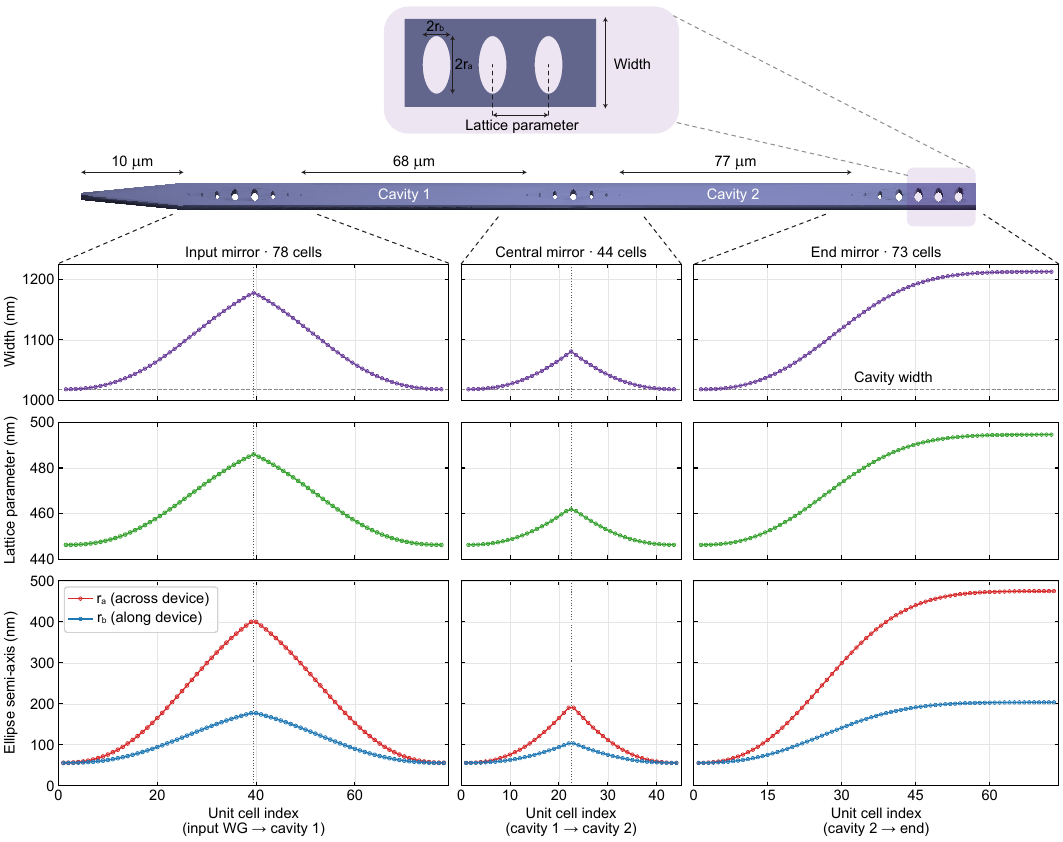}
            \caption{\textbf{Nominal parameters of the coupled Fabry--Pérot resonators.} The plots show the unit-cell parameters (the waveguide width, lattice constant, and ellipse semi-axes, as defined in the schematic on the top of the figure) for the different mirrors forming the coupled cavities, as a function of unit-cell index. The waveguide sections in cavities 1 and 2 have lengths of \SI{68}{\micro\meter} and \SI{77}{\micro\meter}, respectively. The edge coupler is a \SI{10}{\micro\meter}-long tapered waveguide with a \SI{200}{\nano\meter} minimum width which is linearly transitioned into the \SI{1.02}{\micro\meter} width at the start of the input mirror. The thin-film lithium niobate layer is \SI{400}{\nano\meter} thick and is etched to a depth of \SI{300}{\nano\meter}. The resulting sidewall angle is approximately $20^\circ$.
            }
            \label{SIfig:fig1}
        \end{figure}

    The unit-cell parameters of the input, central, and end mirrors are shown in Fig.~\ref{SIfig:fig1}. We use $M = 2.6$ and $\delta = 27.5$, and the values of $p_0$ and $p_\infty$ for each unit-cell parameter are identical for all mirrors. For the end mirror, the parameters reach a plateau for $n > 60$, where they are close to $p_\infty$ and the unit-cell geometry no longer changes appreciably ($< 0.5\,\%$ per cell). In Table~\ref{tab:mirror_taper} we summarize the values of $p_0$ and $p_\infty$ for all the parameters composing our design.

  \begin{table}[htbp]
    \centering
    \caption{Taper parameters of the photonic-crystal mirrors. All four unit-cell parameters share $M = 2.6$ and $\delta = 27.5$.}
    \label{tab:mirror_taper}
    \begin{tabular}{lcc}
        \hline\hline
        Parameter & $p_0$ (nm) & $p_\infty$ (nm) \\
        \hline
        Lattice constant & 446.2 & 494.7 \\
        Semi-axis along the beam, $r_a$ & 55.8 & 203.7 \\
        Semi-axis across the beam, $r_b$ & 55.8 & 475.3 \\
        Beam width & 1018.5 & 1212.5 \\
        \hline\hline
    \end{tabular}
\end{table}

\begin{figure}[t!]
            \centering
            \includegraphics[width=18cm]{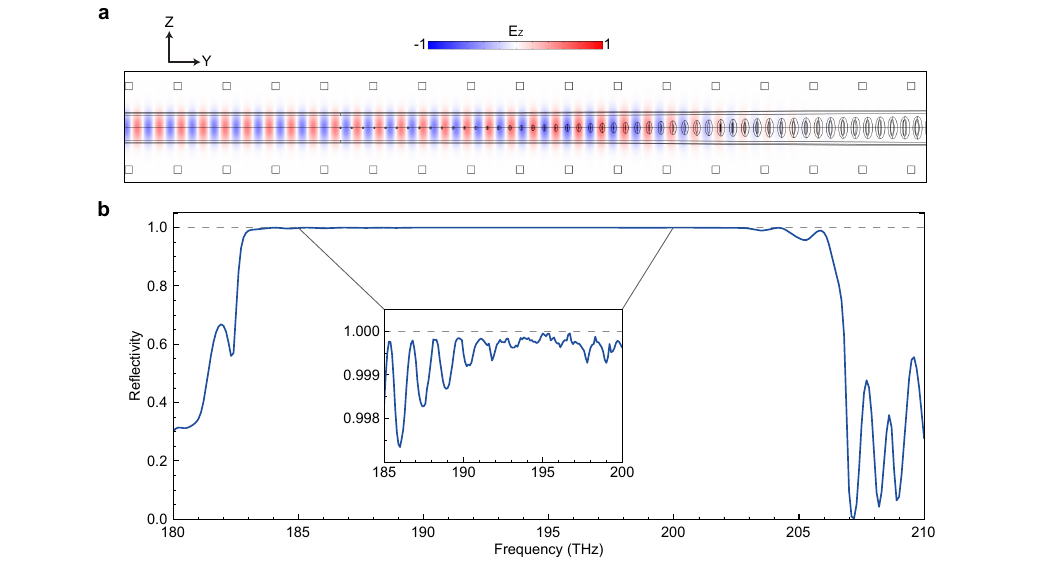}
            \caption{\textbf{Simulated response of the optimized photonic-crystal mirror.} \textbf{a}, Simulated electric-field profile of a Fabry--Pérot cavity, showing the component along the $Z$ crystal direction, for the optimized mirror geometry. The remaining part of the Fabry--Pérot cavity is omitted here to highlight the strong reflectivity at the mirror sections. \textbf{b}, Corresponding reflectivity spectrum, showing a photonic bandgap exceeding $\SI{20}{\THz}$ around the telecom C band. Inset: zoom-in into the 185 - 200 THz band, displaying reflectivities above 0.999.
            }
            \label{SIfig:fig2}
        \end{figure}

    The parameters of our fabricated devices were obtained by optimizing the structure with finite-element method (FEM) simulations in COMSOL Multiphysics using an anisotropic model for TFLN. We first performed unit-cell simulations with periodic boundary conditions to optimize the photonic bandgap of the mirror for the fundamental TE mode, centering it on the telecom C band. The waveguide-mirror adiabatic transition and the positions of the release holes for the device undercut were then adjusted in full-cavity electromagnetic simulations to maximize the optical quality factor. Using perfectly matched layers, these simulations consistently yielded radiation quality factors $Q_\text{opt,rad} > \num{1e7}$.
    Both the outer sidewalls of the waveguide and the inner sidewalls of the elliptical holes were assumed to have an angle of \SI{70}{\degree} with the device plane, extracted via scanning electron microscopy (SEM) images. Figure~\ref{SIfig:fig2}\textbf{a} shows an example of the electric-field component along the crystal Z axis in the resulting Fabry--Pérot cavity. The end mirror reflectivity as a function of frequency is shown in Fig.~\ref{SIfig:fig2}\textbf{b}. This was calculated by launching the fundamental TE mode into the waveguide through an excitation port and computing the power reflected into the same mode. The reflectivity approaches unity over a bandwidth exceeding \SI{20}{\THz}.
    
    \subsection{Quarter-wavelength resonator geometry}
    We designed the mm-wave resonator such that its resonance frequency matches the optical mode splitting, $\omega_+ - \omega_- \approx \SI{260}{\GHz}$. To this end, we performed finite-element simulations in COMSOL Multiphysics in which we varied the lengths of the quarter-wavelength resonator and of its antenna section. The gold electrodes were modeled as a film of thickness \SI{300}{\nano\meter} and conductivity $\sigma_\text{Au} = \SI{4.1e7}{\siemens\per\meter}$. To reduce the computational cost, we omitted the release holes and the waveguides and modeled the TFLN layer as a slab of thickness \SI{100}{\nano\meter}, corresponding to the film that remains after the etch used to pattern our devices. The resonator was placed at the center of a \SI{0.5}{\milli\meter} $\times$ \SI{0.5}{\milli\meter} simulation domain bounded by perfectly matched layers, which capture the radiative losses of the structure. Since the mm-wave fields are strongly influenced by the substrate, we included the full thicknesses of the buried oxide layer (\SI{4.7}{\micro\meter}) and of the silicon handle (\SI{0.5}{\milli\meter}). To capture the frequency shift caused by releasing the devices, the oxide beneath the region where the Fabry--Pérot cavities are in the actual device was replaced by vacuum over a width equal to the undercut measured in our devices.
    
\begin{figure}[t!]
            \centering
            \includegraphics[width=18cm]{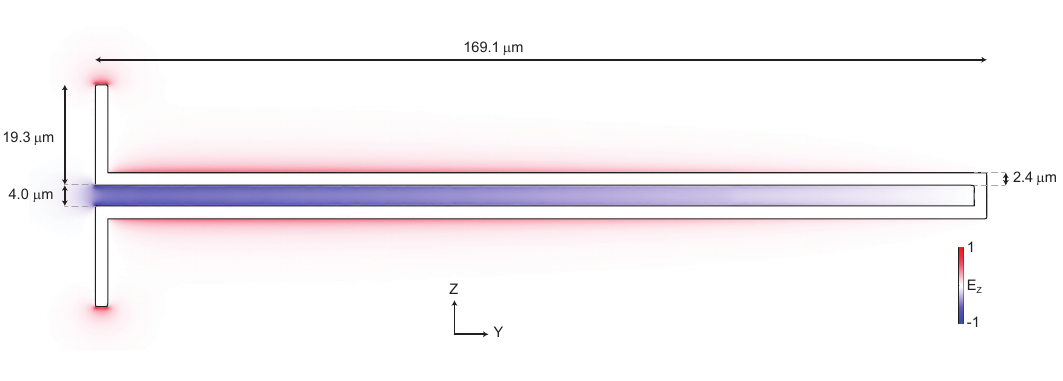}
            \caption{\textbf{Millimeter-wave resonator design.} Sketch of the fabricated device with the relevant geometric parameters and simulated mode profile of the quarter-wavelength resonator, showing the electric-field component along the crystal Z axis. This is the dominant field component and is parallel to the polarization of the fundamental TE mode of the Fabry--Pérot cavities. The simulated mode has a resonance frequency of \SI{264.59}{\GHz} and a quality factor $Q_\text{mmw} = 18$. The release holes and waveguides are omitted from the simulation, and the TFLN layer is modeled as a \SI{100}{\nano\meter}-thick slab.}
            \label{SIfig:fig3}
        \end{figure}

    The relevant parameters of the fabricated devices are sketched and the mode profile of the quarter-wavelength resonator are shown in Figure~\ref{SIfig:fig3}. We display solely the electric-field component along the crystal Z axis, which is the dominant component of the field. Because the fundamental TE mode of the Fabry--Pérot cavities is polarized along the same axis, the two modes interact through the largest electro-optic coefficient of lithium niobate, $r_{33}$, which underlies the large electro-optic coupling rate of our device. The simulations yield a resonance frequency of \SI{264.59}{\GHz} and a quality factor $Q_\text{mmw} \approx 18$, in close agreement with the measured values. We find that, within our model with reduced dimension, $Q_\text{mmw}$ is strongly dominated by conduction losses, $Q_\text{mmw,cond}\approx 19$, rather than by radiation, $Q_\text{mmw,rad}\approx 340$.

\section{Triply-resonant transducer model}
\begin{figure}[t!]
            \centering
            \includegraphics[width=18cm]{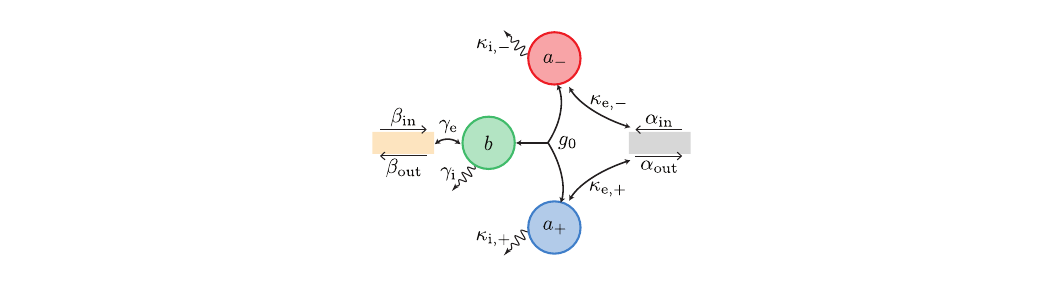}
            \caption{\textbf{Coupled-mode description of the transducer.} The optical modes $a_\pm$ have internal and external loss rates $\kappa_{\text{i},\pm}$ and $\kappa_{\text{e},\pm}$, and are coupled to the same waveguide with input and output amplitudes $\alpha_\text{in}$ and $\alpha_\text{out}$. The millimeter-wave mode $b$ is driven and read out with amplitudes $\beta_\text{in}$ and $\beta_\text{out}$ through an external channel at rate $\gamma_\text{e}$, and dissipates internally at rate $\gamma_\text{i}$. The optical modes are electro-optically coupled to $b$ with a rate $g_0$, which allows coherent conversion between them.}
            \label{SIfig:fig4}
        \end{figure}

 The underlying coupled mode model for triply resonant systems is well documented~\cite{mckenna2020,Han2021}. Here, we lay out the foundations and derive with more detail the case-specific results arising from the reversed dissipation regime, where mm-wave losses are much larger than their optical counterpart. The diagram of the different couplings in the system shown in the main text is reproduced in Fig.~\ref{SIfig:fig4} for convenience.

Assuming that the three interacting modes are well isolated and neglecting off-resonant terms, we write the interaction Hamiltonian of the system as
\begin{equation}
    H = \hbar g_0(a_+ a_-^\dagger b^\dagger + a_+^\dagger a_- b),
    \label{eq:Hamiltonian}
\end{equation}
where $a_+$, $a_-$, and $b$ are the annihilation operators of the modes with frequencies $\omega_+$, $\omega_-$, and $\omega_\text{mmw}$, respectively, as defined in the main text. When mode $a_+$ is coherently driven, linearizing this Hamiltonian yields a two-mode squeezing interaction,
\begin{equation}
    H_\text{TMS} = \hbar g_0\sqrt{n_{\text{p},+}}(a_-^\dagger b^\dagger + a_- b) = \hbar g(a_-^\dagger b^\dagger + a_- b),
    \label{eq:H_TMS}
\end{equation}
where $\sqrt{n_{\text{p},+}}$ is the steady-state coherent amplitude of mode $a_+$, whose phase is set to zero without loss of generality. Analogously, pumping $a_-$ yields a beam-splitter interaction,
\begin{equation}
    H_\text{BS} = \hbar g_0\sqrt{n_{\text{p},-}}(a_+ b^\dagger + a_+^\dagger b) = \hbar g(a_+ b^\dagger + a_+^\dagger b).
    \label{eq:H_BS}
\end{equation}
In both cases, $g = g_0\sqrt{n_{\text{p},\pm}}$ is the pump-enhanced coupling rate.

For simplicity, we now analyze only the beam-splitter interaction, which leads to up-conversion. The linearized Hamiltonian yields the following Heisenberg--Langevin equations of motion for the dynamically relevant modes
\begin{align}
    &\dot a_+ = \left(i\Delta_+ -\kappa_+/2\right)a_+ -i g b -\sqrt{\kappa_{\text{e},+}}\, \delta\alpha_{\text{in,e},+} -\sqrt{\kappa_{\text{i},+}}\, \delta\alpha_{\text{in,i},+},
    \label{eq:Heis_Lang_eq_a}\\
    &\dot b = \left(-i\omega_\text{mmw} -\gamma_\text{mmw}/2\right) b - i g a_+ -\sqrt{\gamma_\text{mmw}}\, \delta\beta_{\text{in}} -\sqrt{\gamma_\text{e}}\,\bar{\beta}_\text{in}e^{-i \omega_\text{d} t},
    \label{eq:Heis_Lang_eq_b}
\end{align}
where $\Delta_+ = \omega_\text{l}-\omega_+$ is the detuning between the laser (at frequency $\omega_\text{l}$) and mode $a_+$ (at frequency $\omega_+$), and $\kappa_+$ is the total loss rate of $a_+$. The operators $\delta\alpha_{\text{in,e},+}$ and $\delta\alpha_{\text{in,i},+}$ describe vacuum fluctuations entering $a_+$ through its external and internal dissipation channels, with coupling rates $\kappa_{\text{e},+}$ and $\kappa_{\text{i},+}$, respectively. Similarly, $\gamma_\text{e}$ and $\gamma_\text{mmw}$ are the external and total loss rates of the mm-wave mode, respectively. The mm-wave mode is driven through its external channel by a coherent input of amplitude $\bar{\beta}_\text{in}$ at frequency $\omega_\text{d}$, while thermal fluctuations $\delta\beta_\text{in}$ enter through all of its dissipation channels. The intracavity pump photon number entering $g$ is $n_{\text{p},-} = 4\kappa_{\text{e},-}|\bar{\alpha}_\text{in}|^2/\kappa_-^2$ for a resonantly pumped mode $a_-$, where $|\bar{\alpha}_\text{in}|^2$ is the input photon flux, and $\kappa_{\text{e},-}$ and $\kappa_-$ are the external coupling rate of $a_-$ to the waveguide and its total loss rate, respectively. Therefore, the effective electro-optic coupling $g$ is dependent on input power and can be tuned.

The noise operators $\delta\alpha_{\text{in,e},+}$, $\delta\alpha_{\text{in,i},+}$, and $\delta\beta_\text{in}$ arise from a Markovian treatment of the baths coupled to our modes, which we assume to be mutually independent~\cite{Clerk2010}. In the frequency domain, the mm-wave noise operators satisfy
\begin{align}
    \langle\delta \beta_\text{in}^\dagger (\omega)\, \delta \beta_\text{in} (\omega^\prime) \rangle &= 2 \pi n_\text{th}\, \delta(\omega+\omega^\prime),
    \label{eq:corr_mmw_a}\\
    \langle\delta \beta_\text{in} (\omega)\, \delta \beta_\text{in}^\dagger (\omega^\prime) \rangle &= 2 \pi (n_\text{th} +1)\, \delta(\omega+\omega^\prime),
    \label{eq:corr_mmw_b}
\end{align}
where $n_\text{th}$ is the thermal occupation of the mm-wave bath. For the optical modes, we assume zero-temperature baths, such that
\begin{align}
    \langle\delta \alpha_{\text{in},k,+}^\dagger (\omega)\, \delta \alpha_{\text{in},k^\prime,+} (\omega^\prime) \rangle &= 0,
    \label{eq:corr_opt_a}\\
    \langle\delta \alpha_{\text{in},k,+} (\omega)\, \delta \alpha_{\text{in},k^\prime,+}^\dagger (\omega^\prime) \rangle &= 2 \pi \delta_{k,k^\prime}\, \delta(\omega+\omega^\prime),
    \label{eq:corr_opt_b}
\end{align}
with $k, k^\prime \in \{\text{e}, \text{i}\}$ and $\delta_{k,k^\prime}$ the Kronecker delta.

\subsection{Thermal sideband analysis}
We now use input-output relations to describe the optical signal generated by the transduction of thermal mm-wave photons. We drop the coherent mm-wave drive in Eqs.~\ref{eq:Heis_Lang_eq_a} and \ref{eq:Heis_Lang_eq_b} and solve them in the frequency domain, using the convention $o(\omega) = \int dt\, e^{i\omega t} o(t)$. The optical amplitude is then
\begin{equation}
    a_+(\omega) = \frac{i g\sqrt{\gamma_\text{mmw}}\,\chi_+(\omega)\chi_\text{mmw}(\omega)}{1 + g^2\chi_+(\omega)\chi_\text{mmw}(\omega)}\,\delta\beta_\text{in}(\omega) + \left(\text{optical noise terms}\right),
    \label{eq:a_plus_omega}
\end{equation}
where we have introduced the optical and mm-wave susceptibilities
\begin{equation}
    \chi_+^{-1}(\omega) = - i\left(\omega + \Delta_+\right)+\frac{\kappa_+}{2}, \qquad
    \chi_\text{mmw}^{-1}(\omega) = - i\left(\omega - \omega_\text{mmw}\right)+\frac{\gamma_\text{mmw}}{2}.
    \label{eq:susceptibilities}
\end{equation}
The term $g^2\chi_+\chi_\text{mmw}$ in the denominator accounts for the back-action of the mm-wave mode on the optical mode.

The field leaving the device through the bus waveguide follows from the input-output relation $\delta\alpha_{\text{out},+} = \delta\alpha_{\text{in,e},+} + \sqrt{\kappa_{\text{e},+}}\,a_+$~\cite{garrison2008}. This gives
\begin{equation}
    \delta\alpha_{\text{out},+}(\omega) = T(\omega)\,\delta\beta_\text{in}(\omega) + \left(\text{optical noise terms}\right), \qquad
    T(\omega) = \frac{i g\sqrt{\kappa_{\text{e},+}\gamma_\text{mmw}}\,\chi_+(\omega)\chi_\text{mmw}(\omega)}{1 + g^2\chi_+(\omega)\chi_\text{mmw}(\omega)}.
    \label{eq:transmission}
\end{equation}

\begin{figure}[t!]
            \centering
            \includegraphics[width=18cm]{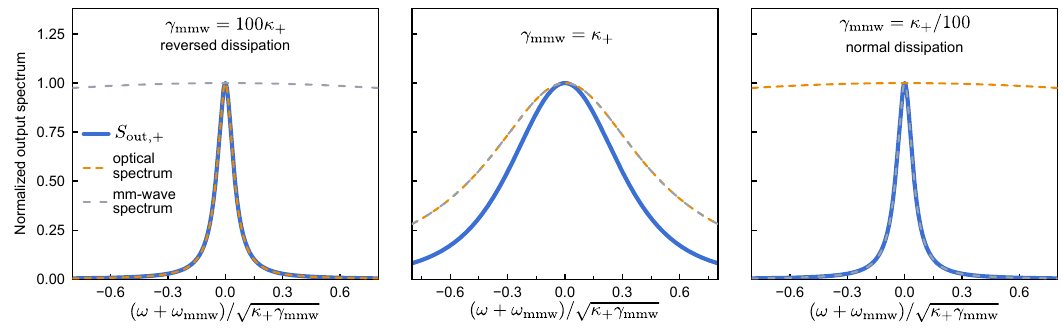}
            \caption{\textbf{Output spectrum for different dissipation regimes.} In the small-$g$ limit, $S_{\text{out},+}$ of Eq.~\ref{eq:S_out_plus} factorizes into the product of the optical and mm-wave Lorentzian responses, so the transduced sideband must fall within both resonances. Deep in the reversed dissipation regime (left panel), the output bandwidth is therefore reduced to the optical linewidth, while in the normal dissipation limit (right panel) it approaches the mm-wave linewidth. When the two linewidths are matched (center panel), both factors contribute equally and the output is narrower than either response, with a full width half maximum = $ \approx 0.64\,\kappa_+$. In all panels the product $\gamma_\text{mmw}\kappa_+$ is held constant and $g$ is chosen such that $C^\text{up} = 0.01 \ll 1$. Together, these results show that the transduction bandwidth is set by the narrower of the two modes and never exceeds it.}
            \label{SIfig:fig5}
        \end{figure}

We define the power spectral density of the output optical field as
\begin{equation}
    S_{\text{out},+}(\omega) = \frac{1}{2\pi}\int^\infty_{-\infty} \langle \delta\alpha^\dagger_{\text{out},+} (\omega)\, \delta\alpha_{\text{out},+} (\omega^\prime) \rangle\, d\omega^\prime,
    \label{eq:PSD_def}
\end{equation}
normalized such that its integral over frequency yields the mean output photon flux in the $a_+$ mode, $\langle \dot{N}_{\text{out},+}\rangle = \frac{1}{2\pi} \int^\infty_{-\infty} S_{\text{out},+}(\omega)\, d\omega$. Since the optical baths are at zero temperature, the optical noise terms do not contribute to this spectrum, which is normally ordered, and we retain only the thermal mm-wave contribution. Our goal is to compute the sideband power emitted by the device due to thermal mm-wave photons, $P^\text{th}_{\text{SB},+} = \hbar\omega_+ \langle \dot{N}_{\text{out},+}\rangle$. Using $\delta\alpha^\dagger_{\text{out},+}(\omega) = T^*(-\omega)\,\delta\beta^\dagger_\text{in}(\omega)$, obtained by Fourier transforming $\delta\alpha^\dagger_{\text{out},+}(t)$, and the thermal correlator of $\delta\beta_\text{in}$, we find
\begin{equation}
\begin{split}
    S_{\text{out},+}(\omega) &= \frac{1}{2\pi}\int^\infty_{-\infty} T^*(-\omega)\, T(\omega^\prime)\, \langle \delta\beta^\dagger_\text{in} (\omega)\, \delta\beta_\text{in} (\omega^\prime) \rangle\, d\omega^\prime
    = |T(-\omega)|^2\, n_\text{th} \\
    &= \frac{g^2 \kappa_{\text{e},+}\gamma_\text{mmw}\, n_\text{th}}{\left|\left[\frac{\kappa_+}{2} + i\left(\omega - \Delta_+\right)\right]\left[\frac{\gamma_\text{mmw}}{2} + i\left(\omega + \omega_\text{mmw}\right)\right] + g^2\right|^2}.
\end{split}
\label{eq:S_out_plus}
\end{equation}
With our Fourier convention, $\delta\alpha^\dagger(\omega) = [\delta\alpha(-\omega)]^\dagger$, so $S_{\text{out},+}$ evaluated at $\omega = -\omega_\text{mmw}$ describes light at $\omega_\text{l} + \omega_\text{mmw}$. Plots for $S_{\text{out},+}(\omega)$ in three different regimes ($\gamma_\text{mmw}\gg\kappa_+$, $\gamma_\text{mmw}=\kappa_+$, and $\gamma_\text{mmw}\ll\kappa_+$) are shown in Fig.~\ref{SIfig:fig5}.

We now assume triple resonance, $\Delta_+ = -\omega_\text{mmw}$, and the reversed-dissipation regime, $\gamma_\text{mmw} \gg \kappa_\pm$. In this case, $S_{\text{out},+}(\omega)$ is sharply peaked around $\omega = -\omega_\text{mmw}$, with a width set by the optical linewidth and hence much smaller than $\gamma_\text{mmw}$. Over this frequency range the mm-wave factor in the denominator of Eq.~\ref{eq:S_out_plus} is nearly constant, $\frac{\gamma_\text{mmw}}{2} + i\left(\omega + \omega_\text{mmw}\right) \approx \frac{\gamma_\text{mmw}}{2}$, which yields
\begin{equation}
    S_{\text{out},+}(\omega) \approx \frac{4 g^2 \kappa_{\text{e},+} n_\text{th}}{\gamma_\text{mmw}\left|\frac{\kappa_+}{2} + \frac{2g^2}{\gamma_\text{mmw}} + i\left(\omega + \omega_\text{mmw}\right)\right|^2}
    = \frac{4 g^2 \kappa_{\text{e},+} n_\text{th}/\gamma_\text{mmw}}{\frac{1}{4}\kappa_+^2\left(1 + C^\text{up}\right)^2 + \left(\omega +\omega_\text{mmw}\right)^2},
    \label{eq:Sout_plus_simplified}
\end{equation}
where $C^\text{up} = 4g^2/(\kappa_+\gamma_\text{mmw})$ is the up-conversion cooperativity defined in the main text, and depends on the input power via $g$. Equation~\ref{eq:Sout_plus_simplified} is a Lorentzian with a full width at half maximum of $\kappa_+(1+C^\text{up})$, which at low cooperativities reduces to $\kappa_+$, as shown in the left panel of Fig.~\ref{SIfig:fig5}, which was obtained from a direct evaluation of Eq.~\ref{eq:S_out_plus}. This broadening is a consequence of dynamical back-action because of the beam-splitter electro-optic interaction with the strongly damped mm-wave resonator, as discussed in the next subsection. Integrating over frequency, we obtain
\begin{equation}
    \langle \dot{N}_{\text{out},+}\rangle \approx \kappa_{\text{e},+}\frac{C^\text{up}}{1+C^\text{up}}\,n_\text{th},
    \qquad
    P^\text{th}_{\text{SB},+} \approx \hbar\omega_+\,\kappa_{\text{e},+}\frac{C^\text{up}}{1+C^\text{up}}\,n_\text{th}.
    \label{eq:Ndot_up}
\end{equation}

We now turn to the two-mode-squeezing interaction, which leads to down-conversion. Pumping $a_+$ and writing $a_-$ in the frame rotating at the pump frequency, the linearized Hamiltonian $H_\text{TMS}$ in Eq.~\ref{eq:H_TMS} yields the coupled-mode equations
\begin{align}
    &\dot a_- = \left(i\Delta_- -\kappa_-/2\right)a_- -i g b^\dagger -\sqrt{\kappa_{\text{e},-}}\, \delta\alpha_{\text{in,e},-} -\sqrt{\kappa_{\text{i},-}}\, \delta\alpha_{\text{in,i},-},
    \label{eq:Heis_Lang_eq_a_TMS}\\
    &\dot b = \left(-i\omega_\text{mmw} -\gamma_\text{mmw}/2\right) b - i g a_-^\dagger -\sqrt{\gamma_\text{mmw}}\, \delta\beta_{\text{in}} -\sqrt{\gamma_\text{e}}\,\bar{\beta}_\text{in}e^{-i \omega_\text{d} t},
    \label{eq:Heis_Lang_eq_b_TMS}
\end{align}
where $g = g_0\sqrt{n_{\text{p},+}}$. The triply resonant condition now reads $\Delta_- = \omega_\text{l}-\omega_- = +\omega_\text{mmw}$, as opposed to $\Delta_+ = -\omega_\text{mmw}$ for up-conversion.

Since $a_-$ couples to $b^\dagger$, the sign of the $g^2$ term in the denominator of Eqs.~\ref{eq:S_out_plus} and \ref{eq:Sout_plus_simplified} is reversed, and the output spectrum $S_{\text{out},-}(\omega) = \frac{1}{2\pi}\int^\infty_{-\infty} \langle \delta\alpha^\dagger_{\text{out},-} (\omega)\, \delta\alpha_{\text{out},-} (\omega^\prime) \rangle\, d\omega^\prime$ involves the anti-normally ordered correlator $\langle\delta\beta_\text{in}(\omega)\,\delta\beta_\text{in}^\dagger(\omega^\prime)\rangle$, which replaces $n_\text{th}$ with $n_\text{th}+1$ and peaks at $\omega = \omega_\text{mmw}$. The resulting output spectrum in the reversed dissipation regime is

\begin{equation}
    S_{\text{out},-}(\omega) \approx \frac{4g^{2}\kappa_{\text{e},-}\left(n_{\text{th}}+1\right)}{\gamma_{\text{mmw}}\left|\dfrac{\kappa_{-}}{2}-\dfrac{2g^{2}}{\gamma_{\text{mmw}}}+i\left(\omega-\omega_{\text{mmw}}\right)\right|^{2}}=\frac{4g^{2}\kappa_{\text{e},-}\left(n_{\text{th}}+1\right)/\gamma_{\text{mmw}}}{\tfrac{1}{4}\kappa_{-}^{2}\left(1-C^{\text{down}}\right)^{2}+\left(\omega-\omega_{\text{mmw}}\right)^{2}},
\end{equation}
with output photon flux and sideband power given by
\begin{equation}
    \langle \dot{N}_{\text{out},-}\rangle \approx \kappa_{\text{e},-}\frac{C^\text{down}}{1-C^\text{down}}\left(n_\text{th}+1\right),
    \qquad
    P^\text{th}_{\text{SB},-} \approx \hbar\omega_-\,\kappa_{\text{e},-}\frac{C^\text{down}}{1-C^\text{down}}\left(n_\text{th}+1\right),
    \label{eq:Ndot_down}
\end{equation}
where $C^\text{down} = 4 g^2/(\kappa_-\gamma_\text{mmw})$. In this case, the two-mode-squeezing interaction narrows the optical linewidth to $\kappa_-(1-C^\text{down})$, and the emitted flux diverges as $C^\text{down}\to 1$, which marks the onset of parametric oscillation, and, as a consequence, Eq.~\ref{eq:Ndot_down} is only valid for $C^\text{down} < 1$. The additional unit in $n_\text{th}+1$ originates from the vacuum fluctuations of the mm-wave mode, such that the down-converted sideband exists even at zero mm-wave temperature, i.e., the process of spontaneous parametric down-conversion.

\subsection{Coherent transduction and back-action}

We now describe how coherently driving the mm-wave mode probes the optical lineshape dressed by the electro-optic interaction. We consider the coherent mm-wave input at frequency $\omega_\text{d}$, neglect the fluctuation terms in Eqs.~\ref{eq:Heis_Lang_eq_a} and \ref{eq:Heis_Lang_eq_b}, and compute the steady-state solutions oscillating at the drive frequency, $b(t) = \bar{b}\,e^{-i\omega_\text{d}t}$ and $a_+(t) = \bar{a}_+e^{-i\omega_\text{d}t}$. Since $a_+$ is defined in the frame rotating at the pump frequency $\omega_\text{l}$, the steady-state solution corresponds to an optical tone at $\omega_\text{l}+\omega_\text{d}$. The equations of motion then become
\begin{align}
    &0 = \left[i\left(\Delta_+ + \omega_\text{d}\right) - \frac{\kappa_+}{2}\right]\bar{a}_+ - i g \bar{b},
    \label{eq:Heis_Lang_eq_a_ss}\\
    &0 = \left[-i\left(\omega_\text{mmw} - \omega_\text{d}\right) - \frac{\gamma_\text{mmw}}{2}\right]\bar{b} - i g \bar{a}_+ - \sqrt{\gamma_\text{e}}\,\bar{\beta}_\text{in}.
    \label{eq:Heis_Lang_eq_b_ss}
\end{align}
Solving Eq.~\ref{eq:Heis_Lang_eq_b_ss} for the mm-wave amplitude gives
\begin{equation}
    \bar{b} = -\chi_\text{mmw}(\omega_\text{d})\left[ig\bar{a}_+ + \sqrt{\gamma_\text{e}}\,\bar{\beta}_\text{in}\right],
    \label{eq:b_ss}
\end{equation}
and substituting it into Eq.~\ref{eq:Heis_Lang_eq_a_ss} yields
\begin{equation}
    \left[\chi_+^{-1}(\omega_\text{d}) + g^2\chi_\text{mmw}(\omega_\text{d})\right]\bar{a}_+ = ig\sqrt{\gamma_\text{e}}\,\chi_\text{mmw}(\omega_\text{d})\,\bar{\beta}_\text{in}.
    \label{eq:a_ss_implicit}
\end{equation}
The steady-state amplitudes are therefore
\begin{equation}
    \bar{a}_+ = ig\sqrt{\gamma_\text{e}}\,\chi_{+}^\text{eff}(\omega_\text{d})\,\chi_\text{mmw}(\omega_\text{d})\,\bar{\beta}_\text{in},
    \qquad
    \bar{b} = -\frac{\sqrt{\gamma_\text{e}}\,\chi_\text{mmw}(\omega_\text{d})}{1 + g^2\chi_+(\omega_\text{d})\chi_\text{mmw}(\omega_\text{d})}\,\bar{\beta}_\text{in},
    \label{eq:ab_ss}
\end{equation}
where we have defined the dressed optical susceptibility
\begin{equation}
    \left[\chi_{+}^\text{eff}(\omega)\right]^{-1} = \chi_+^{-1}(\omega) + g^2\chi_\text{mmw}(\omega).
    \label{eq:chi_eff}
\end{equation}
The electro-optic interaction thus adds a so-called self-energy $g^2\chi_\text{mmw}(\omega)$ to the bare optical response. Collecting its real and imaginary parts, we get
\begin{equation}
    \left[\chi_{+}^\text{eff}(\omega)\right]^{-1} = - i\left[\omega + \Delta_+ - \delta\omega_+(\omega)\right] + \frac{\kappa_+^\text{eff}(\omega)}{2},
    \label{eq:chi_eff_split}
\end{equation}
with
\begin{equation}
    \kappa_+^\text{eff}(\omega) = \kappa_+ + \frac{g^2\gamma_\text{mmw}}{\left(\gamma_\text{mmw}/2\right)^2 + \left(\omega - \omega_\text{mmw}\right)^2},
    \qquad
    \delta\omega_+(\omega) = \frac{g^2\left(\omega - \omega_\text{mmw}\right)}{\left(\gamma_\text{mmw}/2\right)^2 + \left(\omega - \omega_\text{mmw}\right)^2}.
    \label{eq:kappa_eff}
\end{equation}
The real part of the self-energy broadens the optical linewidth, while the imaginary part shifts the optical resonance (the electro-optic spring effect, analogous to the optical spring effect in optomechanics~\cite{Aspelmeyer2014}). At the triply resonant condition, $\Delta_+ = -\omega_\text{mmw}$, the optical response is centered at $\omega = \omega_\text{mmw}$, and in the reversed-dissipation regime the relevant frequencies satisfy $|\omega - \omega_\text{mmw}| \ll \gamma_\text{mmw}$. There, $\kappa_+^\text{eff} \approx \kappa_+ + 4g^2/\gamma_\text{mmw} = \kappa_+(1 + C^\text{up})$, whereas $\delta\omega_+ \approx C^\text{up}\kappa_+(\omega - \omega_\text{mmw})/\gamma_\text{mmw}$ is negligible compared to $\omega - \omega_\text{mmw}$. These approximations are valid in the weak-coupling limit where $2g\ll \gamma_\text{mmw}$. The dressed optical susceptibility therefore reduces to
\begin{equation}
    \left[\chi_{+}^\text{eff}(\omega)\right]^{-1} \approx - i\left(\omega - \omega_\text{mmw}\right) + \frac{\kappa_+\left(1 + C^\text{up}\right)}{2},
    \label{eq:chi_eff_approx}
\end{equation}
i.e., the optical mode responds as a Lorentzian resonance whose linewidth is broadened from $\kappa_+$ to $\kappa_+(1+C^\text{up})$.

Since there is no coherent optical input at $\omega_\text{l} + \omega_\text{d}$, the input--output relation gives $\bar{\alpha}_{\text{out},+} = \sqrt{\kappa_{\text{e},+}}\,\bar{a}_+$, and the output photon flux in the upper sideband is
\begin{equation}
    |\bar{\alpha}_{\text{out},+}|^2 = \eta(\omega_\text{d})\,|\bar{\beta}_\text{in}|^2,
    \qquad
    \eta(\omega_\text{d}) = g^2\kappa_{\text{e},+}\gamma_\text{e}\,|\chi_\text{mmw}(\omega_\text{d})|^2\,|\chi_{+}^\text{eff}(\omega_\text{d})|^2 = \frac{\gamma_\text{e}}{\gamma_\text{mmw}}\,|T(\omega_\text{d})|^2,
    \label{eq:eta_coh}
\end{equation}
where $\eta(\omega_\text{d})$ is the photon-number conversion efficiency and $T(\omega)$ is given in Eq.~\ref{eq:transmission}. The sideband power
\begin{equation}
 P^\text{coh}_{\text{SB},+}(\omega_\text{d}) = \hbar(\omega_\text{l}+\omega_\text{d})\,|\bar{\alpha}_{\text{out},+}|^2 \approx \hbar\omega_+\,\eta(\omega_\text{d})\,|\bar{\beta}_\text{in}|^2
 \label{eq:PSB_plus_coh}
\end{equation}
follows $\eta(\omega_\text{d})$ as the drive frequency is swept. Using Eq.~\ref{eq:chi_eff_approx} together with $|\chi_\text{mmw}(\omega_\text{d})|^2 \approx 4/\gamma_\text{mmw}^2$, we obtain
\begin{equation}
    \eta(\omega_\text{d}) \approx \frac{4g^2\kappa_{\text{e},+}\gamma_\text{e}/\gamma_\text{mmw}^2}{\frac{1}{4}\kappa_+^2\left(1 + C^\text{up}\right)^2 + \left(\omega_\text{d} - \omega_\text{mmw}\right)^2}
    = \frac{\gamma_\text{e}}{\gamma_\text{mmw}}\,\frac{\kappa_{\text{e},+}}{\kappa_+}\,\frac{4C^\text{up}}{\left(1 + C^\text{up}\right)^2}\,\left[\frac{\kappa_+^2\left(1 + C^\text{up}\right)^2/4}{\kappa_+^2\left(1 + C^\text{up}\right)^2/4 + \left(\omega_\text{d} - \omega_\text{mmw}\right)^2}\right],
    \label{eq:eta_lorentzian}
\end{equation}
with peak efficiency at $\omega_\text{d} =  \omega_\text{mmw}$, given by $\eta =  (\gamma_\text{e}/\gamma_\text{mmw})(\kappa_{\text{e},+}/\kappa_+)4C^\text{up}/{(1+C^\text{up})^2}$~\cite{Han2021}. From Eq.~\ref{eq:eta_lorentzian}, sweeping the drive frequency traces a Lorentzian with a full width at half maximum of $\kappa_+(1+C^\text{up})$ (the term in brackets), which is exactly the dressed optical linewidth of Eq.~\ref{eq:chi_eff_approx}. This is also the lineshape of the thermal sideband, Eq.~\ref{eq:Sout_plus_simplified}.

The same procedure applies to down-conversion, where $a_-$ couples to $b^\dagger$ and the converted tone appears at $\omega_\text{l} - \omega_\text{d}$. The two-mode squeezing interaction reverses the sign of the self-energy, $\left[\chi_-^\text{eff}(-\omega_\text{d})\right]^{-1} = \chi_-^{-1}(-\omega_\text{d}) - g^2\chi_\text{mmw}^*(\omega_\text{d})$, which narrows the optical linewidth to $\kappa_-(1 - C^\text{down})$. The resulting efficiency is again a Lorentzian in $\omega_\text{d}$, now with a full width at half maximum of $\kappa_-(1 - C^\text{down})$ and a peak value $(\gamma_\text{e}/\gamma_\text{mmw})(\kappa_{\text{e},-}/\kappa_-)\,4C^\text{down}/(1 - C^\text{down})^2$.

Above $C^\text{down} = 1$, the system undergoes parametric oscillation and the $a_-$ and $b$ modes oscillate simultaneously. Pump depletion limits the gain at its threshold value, creating an electro-optic analog of Brillouin lasing~\cite{Li2012}, with the strongly damped mm-wave mode taking the role of the acoustic wave, usually sitting at frequencies of a few GHz. Moreover, because the two-mode-squeezing interaction generates photon pairs, each photon emitted into the optical sideband is accompanied by a mm-wave photon, of which a fraction $\gamma_\text{e}/\gamma_\text{mmw}$ leaves the resonator through its external port. In this regime, our device therefore also acts as a coherent mm-wave source with an output power
\begin{equation}
    P_\text{mmw,out} = \frac{\omega_\text{mmw}}{\omega_-}\,\frac{\gamma_\text{e}}{\gamma_\text{mmw}}\,\frac{\kappa_-}{\kappa_{\text{e},-}}\,P_{\text{SB},-}.
    \label{eq:P_mmw_out}
\end{equation}
This is a distinctive advantage over Brillouin lasers, in which the partner of each Stokes photon is an acoustic phonon that cannot be directly extracted as an electromagnetic signal. Together with coherent up-conversion below threshold, this makes our device a versatile interface for both detecting and generating mm-wave signals, particularly in applications where free-space or wired delivery is challenging.

\subsection{Calibration of the external mm-wave coupling rate}

Equations~\ref{eq:Ndot_up} and~\ref{eq:PSB_plus_coh} provide a straightforward way to characterize the external coupling of the mm-wave mode, $\gamma_\text{e}$. In our experiments, the thermal and coherent sideband data were acquired in an interleaved fashion, such that the fiber-to-chip coupling, the optical pump power, and the laser--cavity detuning were similar for both. In addition, the mm-wave power $P_\text{mmw}$ incident on the chip was calibrated as described in the main text, such that $|\bar{\beta}_\text{in}|^2 = P_\text{mmw}/(\hbar \omega_\text{mmw})$. Finally, our measurements were performed with an optical spectrum analyzer whose resolution bandwidth ($\approx \SI{3}{\GHz}$) is much larger than the dressed optical linewidths, $\kappa_+(1+C^\text{up})/2\pi$ and $\kappa_-(1-C^\text{down})/2\pi$. The analyzer therefore integrates the full thermal sideband, and the measured sideband powers are described by Eqs.~\ref{eq:Ndot_up} and~\ref{eq:Ndot_down}.

The maximum sideband power in the coherent transduction data is reached at triple resonance, $\Delta_+ = -\omega_\text{mmw}$ and $\omega_\text{d} = \omega_\text{mmw}$, where Eq.~\ref{eq:PSB_plus_coh} reduces to
\begin{equation}
    P^\text{coh}_{\text{SB},+}(\omega_\text{mmw}) = \hbar\omega_+\,\frac{\gamma_\text{e}}{\gamma_\text{mmw}}\,\frac{\kappa_{\text{e},+}}{\kappa_+}\,\frac{4C^\text{up}}{\left(1 + C^\text{up}\right)^2}\,\frac{P_\text{mmw}}{\hbar \omega_\text{mmw}}.
    \label{eq:PSB_plus_coh_max}
\end{equation}
Dividing Eq.~\ref{eq:PSB_plus_coh_max} by Eq.~\ref{eq:Ndot_up} yields
\begin{equation}
    \frac{P^\text{coh}_{\text{SB},+}(\omega_\text{mmw})}{P^\text{th}_{\text{SB},+}} = \frac{\gamma_\text{e}}{\gamma_\text{mmw}}\,\frac{4}{\kappa_+ \left(1+C^\text{up}\right)}\,\frac{P_\text{mmw}}{\hbar \omega_\text{mmw}n_\text{th}}.
    \label{eq:PSB_ratio}
\end{equation}
At low optical pump powers, $C^\text{up}\ll 1$, and Eq.~\ref{eq:PSB_ratio} can be rewritten as
\begin{equation}
    \frac{\gamma_\text{e}}{\gamma_\text{mmw}} \approx \frac{P^\text{coh}_{\text{SB},+}(\omega_\text{mmw})}{P^\text{th}_{\text{SB},+}}\,\frac{\kappa_+\hbar \omega_\text{mmw}n_\text{th}}{4P_\text{mmw}}.
    \label{eq:gamma_ratio}
\end{equation}
We use Eq.~\ref{eq:gamma_ratio} to extract $\gamma_\text{e}/\gamma_\text{mmw}$, which requires no knowledge of $g_0$ or the cooperativity. Here, $n_\text{th}$ is the thermal occupation of the mm-wave bath, which we assume to be the same for the internal and external dissipation channels. Since $\gamma_\text{e}/\gamma_\text{mmw}$ does not depend on the optical pump power, the value extracted at low $C^\text{up}$ also holds at higher pump powers. Combined with the measured conversion efficiency $\eta$ and the optical coupling ratio $\kappa_{\text{e},+}/\kappa_+$ (extracted via reflection measurements), it allows us to determine the internal efficiency $4C^\text{up}/(1 + C^\text{up})^2$ of the transducer directly and self-consistently. Moreover, knowledge of $\gamma_\text{mmw}$ yields $\gamma_\text{e}$ itself and hence an estimate of the number of mm-wave photons in our resonators.

\section{Millimeter-wave spectrum fitting}

The transduction spectrum in Fig.~\textbf{2e} of the main text deviates from a single Lorentzian. We attribute the ripples that appear in the spectrum to the coupling of the mm-wave resonator to the electromagnetic modes of the silicon substrate of our chip. In our experiments, the chips are diced to dimensions of $0.6~\text{cm}\times1.5~\text{cm}$, and the silicon handle is $500~\mu\text{m}$ thick with a refractive index $n_\text{Si} = 3.42$. Consequently, it behaves as a set of low-finesse etalons whose modes form a structured radiation bath for the resonator. The mm-wave source illuminates the back of the chip, so the substrate modes are excited first and the quarter-wavelength resonator is driven through them. In this section, we derive a closed-form model for the resulting frequency-dependent damping, resonance shift, and external coupling, describe the fitting procedure, and discuss the range of validity of the model.

\subsection{Self-energy and effective drive from the substrate modes}

We describe the substrate by a set of modes $c_k$ with frequencies $\omega_k$, linewidths $\gamma_k$, and couplings $J_k$ to the resonator mode $b$,
\begin{align}
    \dot b &= \left(-i\omega_0 - \frac{\gamma_0}{2}\right) b - i\sum_k J_k c_k,\\
    \dot c_k &= \left(-i\omega_k - \frac{\gamma_k}{2}\right) c_k - i J_k b - \sqrt{\gamma_{\text{e},k}}\,\beta_\text{in},
\end{align}
where $\gamma_0$ is the damping of the resonator in the absence of the substrate modes, dominated by ohmic loss in the gold, and $\gamma_{\text{e},k}$ is the rate at which substrate mode $k$ exchanges energy with the driving field. Because the source is on the back of the chip, we take the quarter-wavelength resonator to be excited only through the substrate and not directly. Eliminating the substrate modes in the frequency domain, the mm-wave susceptibility of Eq.~\ref{eq:susceptibilities} becomes
\begin{equation}
    \chi_\text{mmw}^{-1}(\omega) = -i\left(\omega - \omega_0\right) + \frac{\gamma_0}{2} + \Sigma_\text{sub}(\omega),
    \qquad
    \Sigma_\text{sub}(\omega) = \sum_k \frac{J_k^2}{\gamma_k/2 - i\left(\omega - \omega_k\right)},
    \label{eq:sigma_sub}
\end{equation}
which has the same structure as the electro-optic self-energies of Eqs.~\ref{eq:chi_eff}. Writing
\begin{equation}
    \chi_\text{mmw}^{-1}(\omega) = -i\left[\omega - \omega_0 - \delta\omega_\text{sub}(\omega)\right] + \frac{\Gamma_\text{mmw}(\omega)}{2},
\end{equation}
the substrate adds a frequency-dependent damping, $\Gamma_\text{mmw}(\omega) = \gamma_0 + 2\,\text{Re}\,\Sigma_\text{sub}(\omega)$, and shifts the resonance by $\delta\omega_\text{sub}(\omega) = \text{Im}\,\Sigma_\text{sub}(\omega)$. The same elimination also collects the drive terms, giving $\chi_\text{mmw}^{-1}(\omega)\,b = -\sqrt{\gamma_\text{e}(\omega)}\,\beta_\text{in}$ with a frequency-dependent external coupling
\begin{equation}
    \sqrt{\gamma_\text{e}(\omega)} = -i\sum_k \frac{J_k\sqrt{\gamma_{\text{e},k}}}{\gamma_k/2 - i\left(\omega - \omega_k\right)}.
    \label{eq:gamma_e_sum}
\end{equation}
Equations~\ref{eq:sigma_sub} and \ref{eq:gamma_e_sum} have the same form, so a single closed-form evaluation of one of them delivers both.

\subsection{Etalon families and closed-form damping}

We group the substrate modes into $M$ families. Each family corresponds to one etalon geometry, i.e., one bounce path with round-trip delay $\tau_j$. The modes belonging to a family are assumed to have identical linewidths $\gamma_j$, couplings $J_j$, and external couplings $\gamma_{\text{e},j}$. Taking a single $\gamma_j$ for a family is a good approximation because modes from the same family leak out through the same boundaries of the substrate every half a roundtrip, yielding identical loss rates for all of them, as described below. The approximation for $J_j$ assumes that the etalon modes of interest have similar spatial overlap with the quarter-wavelength, which is valid if the interaction happens over a narrow bandwidth; the same argument applies to their overlap with the incident beam, and hence to $\gamma_{\text{e},j}$. These etalon modes form a comb, $\omega_{j,k} = \tilde\omega_j + 2\pi k/\tau_j$. The self-energy of family $j$ follows from Eq.~\ref{eq:sigma_sub} by restricting the sum to its members,
\begin{equation}
    \Sigma_j(\omega) = J_j^2\sum_{k=-\infty}^{\infty}\frac{1}{\gamma_j/2 - i\left(\omega - \omega_{j,k}\right)},
    \qquad
    \omega_{j,k} = \tilde\omega_j + \frac{2\pi k}{\tau_j}.
    \label{eq:sigma_j_sum}
\end{equation}

Because this is a linear system, the contributions of the different families add to form the total substrate self-energy $\Sigma_\text{sub} = \sum^{M}_{j = 1} \Sigma_j$, so adding families is trivial once a closed-form expression is found for the mm-wave spectrum modification induced by one of them.

The real part of Eq.~\ref{eq:sigma_j_sum} is a comb of Lorentzians,
\begin{equation}
    \text{Re}\,\Sigma_j(\omega) = \frac{J_j^2\gamma_j}{2}\sum_{k=-\infty}^{\infty}\frac{1}{\gamma_j^2/4 + \left(\omega - \omega_{j,k}\right)^2},
    \label{eq:re_sigma_j_sum}
\end{equation}
which is most easily evaluated in the time domain. Consistent with the convention $o(\omega) = \int dt\, e^{i\omega t}o(t)$, we define
\begin{equation}
    R_j(t) = \frac{1}{2\pi}\int d\omega\, e^{-i\omega t}\,\text{Re}\,\Sigma_j(\omega).
\end{equation}
Each Lorentzian transforms into a damped oscillation at the mode frequency
\begin{equation}
    \frac{1}{2\pi}\int d\omega\,e^{-i\omega t}\,\frac{\gamma_j/2}{\gamma_j^2/4 + \left(\omega - \omega_{j,k}\right)^2} = \frac{1}{2}\,e^{-i\omega_{j,k}t}\,e^{-\gamma_j|t|/2},
\end{equation}
such that
\begin{equation}
    R_j(t) = \frac{J_j^2}{2}\,e^{-\gamma_j|t|/2}\,e^{-i\tilde\omega_j t}\sum_{k=-\infty}^{\infty}e^{-i2\pi k t/\tau_j}.
\end{equation}
The remaining sum follows from the Poisson summation formula $
    \sum_{k=-\infty}^{\infty} e^{-i2\pi k t/\tau_j} = \tau_j\sum_{m=-\infty}^{\infty}\delta\left(t - m\tau_j\right)$,
and hence each delta function evaluates the terms $e^{-\gamma_j|t|/2}$ and $e^{-i\tilde\omega_j t}$ at $t = m\tau_j$, that is:
\begin{equation}
    R_j(t) = \frac{\bar s_j}{2}\sum_{m=-\infty}^{\infty}\alpha_j^{|m|}\,e^{im\theta_j}\,\delta\left(t - m\tau_j\right),
    \label{eq:echo_train}
\end{equation}
with
\begin{equation}
    \bar s_j = J_j^2\tau_j, \qquad
    \alpha_j = e^{-\gamma_j\tau_j/2}, \qquad
    \theta_j = -\tilde\omega_j\tau_j.
    \label{eq:echo_params}
\end{equation}
In the time domain, the comb of substrate modes is thus a train of echoes at multiples of the round-trip delay $\tau_j$. Since the field of a mode with energy decay rate $\gamma_j$ decays as $e^{-\gamma_j t/2}$, $\alpha_j$ is the amplitude that survives one round trip, and $\theta_j$ fixes the spectral position of the comb.

The imaginary part of $\Sigma_j$ follows from causality or direct calculation. The full expression becomes
\begin{equation}
    \Sigma_j(t) = \bar s_j\left[\frac{\delta(t)}{2} + \sum_{m=1}^{\infty}\alpha_j^{m}\,e^{im\theta_j}\,\delta\left(t - m\tau_j\right)\right].
    \label{eq:causal_echo_train}
\end{equation}
Transforming back to the frequency domain, the sum becomes a geometric series in $z_j = \alpha_j e^{id_j}$, with $d_j(\omega) = \omega\tau_j + \theta_j$,
\begin{equation}
    \Sigma_j(\omega) = \int dt\, e^{i\omega t}\,\Sigma_j(t)
    = \frac{\bar s_j}{2}\left[1 + 2\sum_{m=1}^{\infty}z_j^m\right]
    = \frac{\bar s_j}{2}\,\frac{1 + z_j}{1 - z_j} \equiv \frac{\bar s_j}{2}\,\mathcal{C}(d_j;\alpha_j),
    \label{eq:sigma_j_closed}
\end{equation}
where the series converges since $|z_j| = \alpha_j < 1$. Multiplying the numerator and denominator by $1 - z_j^*$, we obtain
\begin{equation}
    \mathcal{C}(d_j;\alpha_j) = \frac{1 + z_j}{1 - z_j} = \frac{1 - \alpha_j^2 + 2i\alpha_j\sin d_j}{1 - 2\alpha_j\cos d_j + \alpha_j^2} \equiv P(d_j;\alpha_j) + iQ(d_j;\alpha_j),
\end{equation}
with
\begin{equation}
    P(d;\alpha) = \frac{1-\alpha^2}{1 - 2\alpha\cos d + \alpha^2} = 1 + 2\sum_{m=1}^{\infty}\alpha^m\cos(md),
    \qquad
    Q(d;\alpha) = \frac{2\alpha\sin d}{1 - 2\alpha\cos d + \alpha^2} = 2\sum_{m=1}^{\infty}\alpha^m\sin(md).
    \label{eq:PQ}
\end{equation}
Therefore,
\begin{equation}
    2\,\text{Re}\,\Sigma_j(\omega) = \bar s_j\,P(d_j;\alpha_j), \qquad
    \text{Im}\,\Sigma_j(\omega) = \frac{\bar s_j}{2}\,Q(d_j;\alpha_j).
    \label{eq:sigma_j_ReIm}
\end{equation}

The function $P$ is, up to normalization, the Airy function of a Fabry--P\'erot etalon with round-trip amplitude survival $\alpha_j$. It is strictly positive, oscillates between $(1-\alpha)/(1+\alpha)$ and $(1+\alpha)/(1-\alpha)$, and has unit mean over one period, since all the cosine terms vanish when averaging over $d$. Hence, $\bar s_j$ is the mean damping added by family $j$. This is exactly what one expects from Fermi's golden rule when accounting for the mode density of the comb, $\tau_j/2\pi$ per unit angular frequency: $2\pi J_j^2\,(\tau_j/2\pi) = J_j^2\tau_j = \bar{s}_j$~\cite{garrison2008}. In the limit $\alpha_j \rightarrow 0$, i.e., no echoes from reflection, the family acts as a flat continuum that adds a constant damping $\bar s_j$ without shifting the resonance. The ripples in the damping and the associated resonance shifts therefore arise entirely from having these reflections at the chip facets.

Finally, summing the self-energies of the families as $\Sigma_\text{sub} = \sum_j\Sigma_j$, the damping and resonance shift of the mm-wave resonator are
\begin{equation}
    \Gamma_\text{mmw}(\omega) = \gamma_0 + \sum_{j=1}^{M}\bar s_j\,P(d_j;\alpha_j),
    \qquad
    \delta\omega_\text{sub}(\omega) = \sum_{j=1}^{M}\frac{\bar s_j}{2}\,Q(d_j;\alpha_j).
    \label{eq:sub_damping}
\end{equation}
Because $P > 0$ and $\bar s_j \geq 0$, the substrate only adds damping, $\Gamma_\text{mmw}(\omega) > \gamma_0$ at all frequencies.

\subsection{Effective external coupling}

The drive of Eq.~\ref{eq:gamma_e_sum} is the same comb sum with one power of $J_k$ removed, so it closes in the same way. Restricting Eq.~\ref{eq:gamma_e_sum} to family $j$ and comparing with Eq.~\ref{eq:sigma_j_closed}, which carries the extra factor $J_j^2 = \bar s_j/\tau_j$, gives
\begin{equation}
    \sqrt{\gamma_\text{e}(\omega)} = -i\sum_{j=1}^{M}\frac{\tau_j}{2}\,J_j\sqrt{\gamma_{\text{e},j}}\;\mathcal{C}(d_j;\alpha_j),
    \qquad
    \gamma_\text{e}(\omega) = \left|\sum_{j=1}^{M} u_j\,\mathcal{C}(d_j;\alpha_j)\right|^2,
    \qquad
    u_j = \frac{\tau_j}{2}J_j\sqrt{\gamma_{\text{e},j}}.
    \label{eq:gamma_e_closed}
\end{equation}
If we assume that the incident beam overlaps with all the families in a similar way, we can write the external coupling as $\gamma_{\text{e},j} = \Lambda\gamma_j$ with a single geometric factor $\Lambda$ applying to all families. Using $J_j = \sqrt{\bar s_j/\tau_j}$ the coefficients in the sum reduce to $u_j \propto \sqrt{\tau_j\,\bar s_j}$ up to a global constant. The effective drive then introduces no additional parameters, as it is fixed by the same $\tau_j$, $\theta_j$, $\bar s_j$ and $\alpha$ that set the damping and frequency shift.

Using Eqs.~\ref{eq:sub_damping} and \ref{eq:gamma_e_closed}, the output sideband power from the bare waveguides used in the mm-wave lineshape measurements (Fig.~\textbf{2e} of the main text) is
\begin{equation}
    P_\text{SB,wg}(\omega_\text{d}) = A\,\frac{\left|\sum_{j}\sqrt{\tau_j\bar s_j}\;\mathcal{C}(d_j;\alpha_j)\right|^2}{\left[\omega_\text{d} - \omega_0 - \delta\omega_\text{sub}(\omega_\text{d})\right]^2 + \Gamma_\text{mmw}(\omega_\text{d})^2/4},
    \label{eq:P_coh}
\end{equation}
where $A$ absorbs $\Lambda$, the optical response, and the detection efficiency, all assumed constant. This is the expression we use in our fitting process.

Importantly, all transduction results derived in the previous sections assume constant values of $\omega_\text{mmw}$ and $\gamma_\text{mmw}$. For these, we use the frequency averages of the substrate-modified resonance frequency and damping. Since $P$ has unit mean and $Q$ has zero mean over one period, these averages are $\gamma_\text{mmw} = \gamma_0 + \sum_j \bar s_j$ and $\omega_\text{mmw} = \omega_0$. The standard deviation of the oscillations around the mean values (as described by Eq.~\ref{eq:sub_damping}) define the uncertainty intervals reported in the main text. Within these intervals, the measured optical mode splitting, $\omega_+ - \omega_-$, matches $\omega_\text{mmw}$, which justifies treating the system as close to triply resonant. Deviations from the triple resonance regime would lower the transduction efficiency, such that reproducing the measured data would require larger values of $g_0$ than those reported in Table~\ref{tab:params}. The reported electro-optic coupling rates should therefore be regarded as conservative estimates.

\subsection{Fitting procedure}

For $M$ families, the model of Eq.~\ref{eq:P_coh} contains the amplitude $A$, the resonance frequency $\omega_\text{mmw}$, and, for each family, the delay $\tau_j$, the phase $\theta_j$, and the mean damping $\bar s_j$. A single round-trip survival $\alpha$ is shared by all families, giving 12 free parameters for $M = 3$, which is what we use. The intrinsic damping $\gamma_0$ is fixed, as discussed below.

We fit the round-trip survival rather than the family linewidths because $\alpha$ has a simple physical reference value. At normal incidence, the field reflection coefficient of a silicon--air facet is $r_\text{F} = (n_\text{Si}-1)/(n_\text{Si}+1) = 0.55$, such that a round trip between two facets retains an amplitude fraction $\alpha_\text{F} = r_\text{F}^2 = 0.30$, which coincides with the power reflectivity of a single facet. The fitted $\alpha$ is an effective round-trip survival in the sense that, in addition to the facet reflections, it accounts for absorption in the silicon and for diffraction losses along each path, and is therefore expected to fall somewhat below $\alpha_\text{F}$. We bound $\alpha$ to $[0.02, 0.60]$. The fit settles well inside it and does not approach either bound. Importantly, $\gamma_j$ and $Q_j$ can be directly extracted from our analysis as
\begin{equation}
    \gamma_j = -\frac{2\ln\alpha_j}{\tau_j}, \qquad Q_j = \frac{\omega_\text{mmw}\,\tau_j}{-2\ln\alpha_j}.
    \label{eq:Qj}
\end{equation}
Bounding the linewidths instead of $\alpha$ would allow short paths to reach unphysical reflectivities, since $\alpha_j = e^{-\gamma_j\tau_j/2}$ approaches unity for small $\tau_j$. With a shared $\alpha$, $Q_j \propto \tau_j$ holds exactly, which is what we expect (longer etalon paths will have higher $Q$ because signals bounce off the facets less often). We further find that this choice yields a lower fitting residual than one reflectivity per family.

To prevent two families from collapsing onto the same delay, the delays are parametrized as an ordered array, $\tau_1 = \tau_\text{min} + u_1$ and $\tau_j = \tau_{j-1} + \Delta\tau + u_j$, with $u_j \geq 0$ and $\Delta\tau = 1/\Delta f_\text{sweep}$, i.e., the delay resolution or the inverse of the swept mm-wave frequency range. Two ripples whose delays differ by less than $\Delta\tau$ cannot be distinguished in our data. In our results, the fitted delays are separated by at least 63~ps, which exceeds the resolution of the 29.6~GHz fit window, $\approx 34$~ps. The lower bound $\tau_\text{min} = 0.05$~ns corresponds to a period of 20~GHz, below which a ripple is indistinguishable from the envelope, because it becomes too similar to the linewidth of the mm-wave mode. The upper bound, $\tau_\text{max} = 2n_\text{Si}L_\text{diag}/c = 0.37$~ns, is the round trip in bulk silicon across the chip diagonal. The fit is a bounded least-squares minimization of the data residual in which the delays are seeded with the peaks of the Fourier transform of the spectrum after subtracting its Savitzky--Golay envelope. The result is then the ripple on the spectrum. We then perform 14 restarts of the fitting process with random initial phases and keep the overall best solution; increasing this to 60 restarts reproduces the same solution.

The intrinsic damping $\gamma_0$ is poorly constrained by the data, because variations in its value can be compensated by the average damping added by the etalon families $\sum \bar s_j$. We therefore fix $\gamma_0 = \omega_\text{mmw}/Q_\text{cond}$ with $Q_\text{cond} = 20$, i.e., $\gamma_0/2\pi \approx 13.2$~GHz. This value is compatible with conductor-loss estimates for our films and devices ($\sigma_\text{Au} = 4.1\times10^7$~S/m, 300 nm Au film) described before. The values of $\bar s_j$ and $\alpha$ are conditional on this choice, whereas the delays and mean quality factor $\omega_\text{mmw}/(\gamma_0 + \sum\bar s_j)$ are nearly insensitive to it: varying $Q_\text{cond}$ between 20 and 25 changes the fitted delays by less than 3\% and the residual by 2\%.

\subsection{Results}

With $M = 3$ families, the fit yields a resonance frequency $\omega_0/2\pi = 264.37$~GHz, a shared round-trip survival $\alpha = 0.222$, and a residual of $3.3$ times the noise floor of the measurement. Table~\ref{tab:families} lists the fitted families. Each delay is converted into an optical path length, $n_\text{g}d_j = c\tau_j/2$, and assigned to a chip dimension ($0.6$ or 1.5~cm) when the resulting group refractive index lies between 1.5 and $n_\text{Si}$. The fitted $\alpha$ sits below the bare Fresnel value $\alpha_\text{F} = 0.30$, as expected once absorption and diffraction along each path are included, and it is reached without approaching the bounds.

\begin{table}[h]
\centering
\small
\setlength{\tabcolsep}{4pt}
\caption{Fitted substrate-mode families ($M = 3$), ordered by round-trip delay. The round-trip amplitude survival, $\alpha = 0.222$, is shared by all families; $\gamma_j$ and $Q_j$ are derived from Eq.~\ref{eq:Qj}, and $J_j = \sqrt{\bar s_j/\tau_j}$. The intrinsic damping is fixed at $\gamma_0/2\pi = 13.24$~GHz ($Q_\text{cond} = 20$), giving $\gamma_\text{mmw}/2\pi = 14.76$~GHz.}
\label{tab:families}
\begin{tabular}{ccccccccc}
\hline
$j$ & $\tau_j$ (ns) & $1/\tau_j$ (GHz) & $n_\text{g}d_j$ (cm) & assignment (chip dimension, $n_\text{g}$) & $\bar s_j/2\pi$ (GHz) & $\gamma_j/2\pi$ (GHz) & $Q_j$ & $J_j/2\pi$ (GHz) \\
\hline
1 & 0.1017 & 9.83 & 1.52 & 0.6~cm, 2.54 & 0.51 & 4.71 & 56  & 0.90 \\
2 & 0.2131 & 4.69 & 3.19 & 1.5~cm, 2.13 & 0.50 & 2.25 & 118 & 0.61 \\
3 & 0.2764 & 3.62 & 4.14 & 1.5~cm, 2.76 & 0.50 & 1.73 & 153 & 0.54 \\
\hline
\end{tabular}
\end{table}

We can also roughly estimate what the model predicts the external coupling of the mm-wave resonator, which is measured independently in our experiments. Evaluating Eq.~\ref{eq:gamma_e_closed} with the fitted parameters and averaging over the fit window gives $\gamma_\text{e}/\gamma_\text{mmw} = 0.24\Lambda$, where $\Lambda = \gamma_{\text{e},j}/\gamma_j$ is the fraction of a substrate mode's radiative loss that would be collected by our WR-2.8 used to drive the chip. A substrate mode is spread over the whole chip, so we can approximate its radiation to leave the substrate through its full $0.6~\text{cm}\times1.5~\text{cm}$ facet area. The WR-2.8 aperture covers only a small portion of that area, given by $0.711~\text{mm}\times0.356~\text{mm}$. Taking their ratio, we get an estimate of $\Lambda \approx 2.8\times10^{-3}$. The predicted external coupling is then $\gamma_\text{e}/\gamma_\text{mmw} = 6.8\times10^{-4}$, which is within 20\% of the measured $8.5\times10^{-4}$. This is rather a consistency check than a precise determination of the value, but it nonetheless supports the picture in which the resonator is driven through the substrate. A drive coupling directly to the quarter-wavelength resonator would have to be adjusted independently to reproduce this number, whereas here it follows from the same parameters that describe the ripples and a simple overlap estimate.

\bibliographystyle{paper}
\bibliography{bibliography}

\end{document}